# Local magnetic correlations and light-sensitive centers in the $Cr_2AlC$ MAX phase

Małgorzata Wierzbowska[1,*], Anna Basa[2], Kazuhiro Marumoto[3,4,5], Jiaxi Wang[3], Katarzyna Gas[6,7,8], Maciej Sawicki[7], Kacper Sierakowski[1], Karel Carva[9], Michał Boćkowski[1]

[1] Institute of High Pressure Physics Polish Academy of Sciences, Sokolowska 29/37, 01-142 Warsaw, Poland
[2] Department of Chemistry, University of Bialystok, Ciolkowskiego 1K, 15-245 Bialystok
[3]Department of Materials Science, Institute of Pure and Applied Sciences, University of Tsukuba, Tsukuba, Ibaraki 305-8573, Japan
[4] Research Center for Organic-Inorganic Quantum Spin Science and Technology (OIQSST), University of Tsukuba, Tsukuba, Ibaraki 305-8573, Japan
[5] Tsukuba Research Center for Energy Materials Science (TREMS), University of Tsukuba, Ibaraki 305-8571, Japan
[6] Center for Science and Innovation in Spintronics, Tohoku University, 2-1-1 Katahira, Aoba-ku, Sendai 980-8577, Japan
[7] Institute of Physics, Polish Academy of Sciences, Aleja Lotnikow 32/46, PL-02668 Warsaw, Poland
[8] Laboratory for Nanoelectronics and Spintronics, Research Institute of Electrical Communication, Tohoku University, 2-1-1 Katahira, Aoba-ku, Sendai 980-8577, Japan
[9] Charles University, Faculty of Mathematics and Physics, DCMP, Ke Karlovu 5, 12116 Prague 2, Czech Republic

[*] corresponding author – Małgorzata Wierzbowska
e-mail: wierzbowska@unipress.waw.pl, Phone: +48 22 6325010, Fax: +48 22 6324218

## Abstract

$Cr_2AlC$ MAX phase is synthesized by high-pressure solid-state annealing and investigated as a candidate platform for optically responsive magnetism. Structural characterization confirms the formation of the $Cr_2AlC$ phase, while magnetic and optical-magnetic properties are examined by superconducting quantum interference device (SQUID) magnetometry, electron spin resonance (ESR), and *first principles* calculations. SQUID magnetometry identifies $Cr_2AlC$ as a weak, field-linear metallic paramagnet dominated by Pauli-like susceptibility of itinerant Cr-derived states. Its non-monotonic temperature dependence is described by an additional contribution from antiferromagnetically coupled Cr-Cr dimers, whereas the low-temperature Curie-like upturn originates from only a trace population of localized Cr centers. Under red-light illumination, SQUID magnetometry does not reveal an intrinsic macroscopic optomagnetic response. In contrast, ESR at 4 K shows a reversible light-induced reduction of a local magnetic signal, but the optically modified spin population corresponds only to several tens of ppm of the Cr sublattice. *Ab initio* Bethe-Salpeter equation (*ai*-BSE) calculations combined with the maximally localized Wannier function analysis suggest that optical excitation can redistribute spin polarization between neighboring Cr sites with the opposite local moments. The combined experiment-theory approach therefore establishes the hierarchy of magnetic contributions in $Cr_2AlC$ and identifies the microscopic origin of its local optical sensitivity.

## 1. Introduction

Controlled manipulation of magnetization might be utilized to store information. The industry and society has a need to increase the density of information, speed up of the reading and writing processes, and improve energy efficiency. Optical manipulation of magnetism has important advantages over traditional application of magnetic fields and represents significant step towards this goal. This concept has a potential to form the technology for the construction of novel magnetoelectronic and optomagnetic devices acting not only as information storage, but also as spin current generators/injectors, sensors, transducers or spin transistors [1].

The discovery of optical manipulation of magnetization employing femtosecond lasers [2] attracted a lot of interest both from basic research and industry. It represented a source of novel information on magnetism and introduced a possibility to examine magnetic systems far from equilibrium [3]. Multiple microscopic mechanisms were proposed to explain how the interaction with photons in the optical range induced magnetization changes. These included electron-phonon scattering [4–6], electron-magnon scattering [7, 8], spin-dependent transport of excited electrons [9], the optically induced intersite spin transfer in multicomponent systems [10] and many others [11]. Notably, many of these mechanisms strongly relied on the spin-orbit coupling [12], while others did not require it. The mechanisms were supported by multiple arguments, and debates about their role did not yet lead to conclusions. It appeared that more of such mechanisms could be present simultaneously, and for different systems diverse effects and processes dominated [13]. Recent calculations demonstrated difficulties connected to the interpretation of physical observables in the case of strong coupling between electrons and light, and revealed the importance of electron correlations [14].

Particularly interesting problem was the magnetization manipulation in antiferromagnets (AFM) [15]. The presence of antiparallel aligned sublattices introduced a new quality into the problem of optically induced magnetization change since each of these sublattices could act as the source or sink of angular momentum for the other one [16]. On the other hand, an observation of magnetization dynamics in the AFMs was significantly more challenging problem than in the ferromagnetic (FM) systems due to the absence of total magnetization. Therefore, successful measurements of spin dynamics in the AFMs had been made only in the last few years. Evolution of sublattice magnetization had been measured in a Gd-based layered AFM employing the resonant x-ray diffraction (focused on peak connected to the AF order) and ARPES [17, 18]. It was also possible to study the dynamics of AFMs indirectly, by measuring the spin emission induced by the optical pump [19]. Optical pumping led to a change of the resistance in AFM CuMnAs that was associated with the magnetization reorientation [20]. Theoretical works focused on the possibility to rotate the magnetization direction in the AFM materials via a torque exerted by light [21], especially for the case with the circularly polarized light where the inverse Faraday effect took place [22].

MAX phases are ternary transition-metal-based carbides, nitrides and carbonitrides given by formula $M_{n+1}AX_n$, where M represents an early transition metal, A is an A-group element (mostly groups IIIA and IVA), and X denotes carbon or nitrogen [23]. They are precursors to obtain MXenes, materials with great promise for multiple applications, including their use in flexible and wearable electronics [24, 25]. This materials platform is therefore relevant for attempts to combine robust layered conductors with magnetic or spintronic functionality [26, 27]. Notably, MXenes are also known for good stability at ambient conditions [25], as compared *e.g.* to transition metal trihalides exhibiting 2D magnetism [28]. Magnetic order had been predicted in many of MXene type compounds, in some cases persisting even up to room temperature [24, 25], but the related experimental data were scarce. First experimental evidence of magnetic ordering had been found in $Cr_2TiC_2T_x$ [29] ($T_x$ denoted stabilizing surface terminations). The possibility to manipulate magnetism had been demonstrated in $Ti_3C_2T_x$, which exhibited weak ferro- or antiferromagnetism that could be further tuned by doping [30]. Magnetism had also been sought in the MAX phases, and the presence and stability of magnetic order was found to strongly depend on sample quality [23].

In $Cr_2AlC$, the indications of weak magnetism (originating from *ab initio* calculations and from x-ray magnetic circular dichroism) had been heavily debated [31], most works concluded that it was nonmagnetic without extra doping [32–33]. Due to the above given reasons it is highly interesting to examine optical control of magnetism in a material from the MAX family, where this phenomenon has remained largely unexplored experimentally.

Here, we combine experiment and theory to assess the magnetic response and optical switching-off magnetism in the $Cr_2AlC$ MAX phase. We synthesize $Cr_2AlC$ and characterize its structure and composition by x-ray diffraction, electron microscopy and energy-dispersive x-ray spectroscopy. The magnetic response is examined by superconducting quantum interference device (SQUID) magnetometry and electron spin resonance (ESR), combining measurements in the dark and influenced by illumination. Magnetometry identifies $Cr_2AlC$ as a weak metallic paramagnet dominated by a Pauli-like susceptibility, with an additional non-monotonic contribution assigned to local antiferromagnetic Cr-Cr correlations and only a trace Curie-like term from dilute localized centers. No macroscopic light-induced magnetic response is detected by magnetometry under red-light illumination. In contrast, ESR reveals a reversible light-induced reduction of a local magnetic signal at 4 K, which corresponds to only a very small fraction of Cr sites. To interpret this local optical sensitivity, we perform *ab initio* calculations of the electronic structure and optical transitions solving the Bethe-Salpeter equation at zero temperature, combined with an analysis of the band character based on maximally localized Wannier functions. Our results establish the hierarchy of magnetic contributions in $Cr_2AlC$ and identify the local character of the optically sensitive response. This provides a useful reference point for future searches of magnetism and its optical switching in MAX phases at low temperatures. These effects would be also relevant for the related MXene systems in which the surface terminations, defects and reduced dimensionality may enhance magnetic states that remain only marginal in the bulk $Cr_2AlC$.

## 2. Materials and methods

### 2.1. Materials Growth

$Cr_2AlC$ has been obtained by a solid state reaction (SSR). The powders: chromium (100 mesh; 99.5 % Aldrich), aluminium (325 mesh; 99.5 % Alfa Acesar) and carbon (graphite, < 20 μm Aldrich) have been mixed in a Cr:Al:C molar ratio of 2:1 and 1:1, respectively. A 10 % excess of Al has been used due to its low melting point, which leads to a partial metal loss during the thermal process [35,36]. The weighed elements have been placed in grinding containers (bowl and balls made of zirconium oxide) under an argon atmosphere and then ground for 3 h in a high-energy ball mill (Pulverisette 7, Fritsch, Idar-Oberstein, Germany) to obtain a homogeneous mixture with fragmented grains. The grinding has been performed in 12 cycles, with each cycle comprising 15 min of grinding at 300 rpm and 15 min of rest. After grinding, pellets with a diameter of 1 cm have been produced from the powders using a laboratory press with hydraulic pump for tablets (MP150, Maassen GmbH, Germany). The samples have been treated using ultra-high-pressure annealing (UHPA) technology [37]. The pellets have been placed in a graphite crucible and separated by graphite spacers. Annealing has been carried out for 15 hours at 1450 ºC and under an argon pressure of 100 MPa. The heating and cooling rate have been 600 ºC/h. After synthesis, the pellets have been crushed, and the resulting powder has been thoroughly mixed, characterized, and divided into portions for SQUID magnetometry and ESR measurements.

### 2.2. Structural Characterization

Diffraction studies have been performed in order to confirm the synthesis of $Cr_2AlC$ using a SuperNova diffractometer (Rigaku, Yarnton, Oxfordshire, UK) equipped with a CCD detector working with $CuK\alpha_1$ ($\lambda$ = 1.54056 Å) radiation source and operating at 50 kV and 0.8 mA. The sample-to-detector distance has been 150 mm and the diffractograms have been recorded in the range of 10-120 2Theta. The sample images have been taken with a scanning electron microscope (SEM, Inspect S50, FEI Company, Hillsboro, Oregon, USA) equipped with a tungsten source and operating at 10 kV. Transmission electron microscopy (TEM, FEI Company, Teknai T20 G2 X-TWIN, Hillsboro, OR, USA) operating at 200 kV and equipped with a $LaB_6$ source has been also used. The sample has been placed on a 400-mesh copper grid covered with a film of holey carbon. To determine the elemental composition of the samples, the x-ray energy dispersion (EDX) spectroscopy measurement has been performed using an EDAX DPP-II analyzer installed on TEM.

### 2.3. Magnetic measurements

Magnetic measurements have been carried out using a Quantum Design MPMS XL superconducting quantum interference device (SQUID) magnetometer equipped with the reciprocating sample option (RSO) to enhance the signal-to-noise ratio. All experiments have been conducted following established protocols for high-precision and high-sensitivity integral SQUID magnetometry to ensure accurate determination of weak magnetic signals, minimize experimental artifacts, and guarantee full traceability of the measured magnetic response [38].

Measurements are performed over the full temperature range of 2 – 400 K, employing two distinct setups optimized for different thermal conditions. For temperatures up to approximately 340 K, the sample, a powdered material, has been enclosed in a standard polycarbonate capsule placed inside an in-situ compensating sample holder, which significantly suppresses the magnetic background signal originating from the capsule [39]. Specifically, the signal from an empty capsule with a typical mass of about 34 mg reduces to the equivalent of approximately 1 mg, yielding a background suppression factor of ~30, making it small in a comparison to a response of the powdered sample, which amounts to 192 mg, significantly simplifying the analysis.

Illumination-dependent magnetometry is performed using a custom prepared setup. All relevant technical details, including the holder geometry, sample preparation, wiring, measurement mode and thermal stability tests, are provided in the Supplementary Information (SI), Sec. S1 (including references [40-44]).

### 2.4. ESR Measurements

The $Cr_2AlC$ samples for the ESR measurements have been prepared as follows: A sample of a measured amount (1 mg) has been bounded to a polyethylene terephthalate (PET) substrate using adhesive. The PET substrate with the sample has been then placed in a sample tube. After evacuating the air, the tube has been filled with helium gas and sealed. ESR measurements have been taken using an X-band ESR spectrometer (JES-FA200; JEOL) at room temperature and 4 K. The $g$-factor, the line width between a spectral peak and valley ($\Delta H_{pp}$), and the number of spins ($N_{spin}$) have been calibrated based on the signals of a $Mn^{2+}$ standard sample. The measurement sample has been placed in the center of the ESR cavity resonator, and the $Mn^{2+}$ standard sample has been placed on the inner wall of the same ESR cavity resonator. The measurement and standard samples have been spatially separated and measured simultaneously in the same ESR cavity resonator. The temperature of the measurement sample has been variable. On the other hand, the temperature of the $Mn^{2+}$ standard sample has been room temperature. The $N_{spin}$ of the $Mn^{2+}$ marker sample has been calibrated in advance using another standard sample 4-hydroxy-2,2,6,6-tetramethylpiperidin-1-oxyl (TEMPOL) solution [45,46]. Simulated solar irradiation (AM1.5G) with a 100 mW $cm^{-2}$ intensity has been used for the ESR measurements under light irradiation using a solar simulator (OTENTOSUN-150LX; Bunkoukeiki Co., Ltd.). The scheme of the ESR setup is presented in Figure 1.

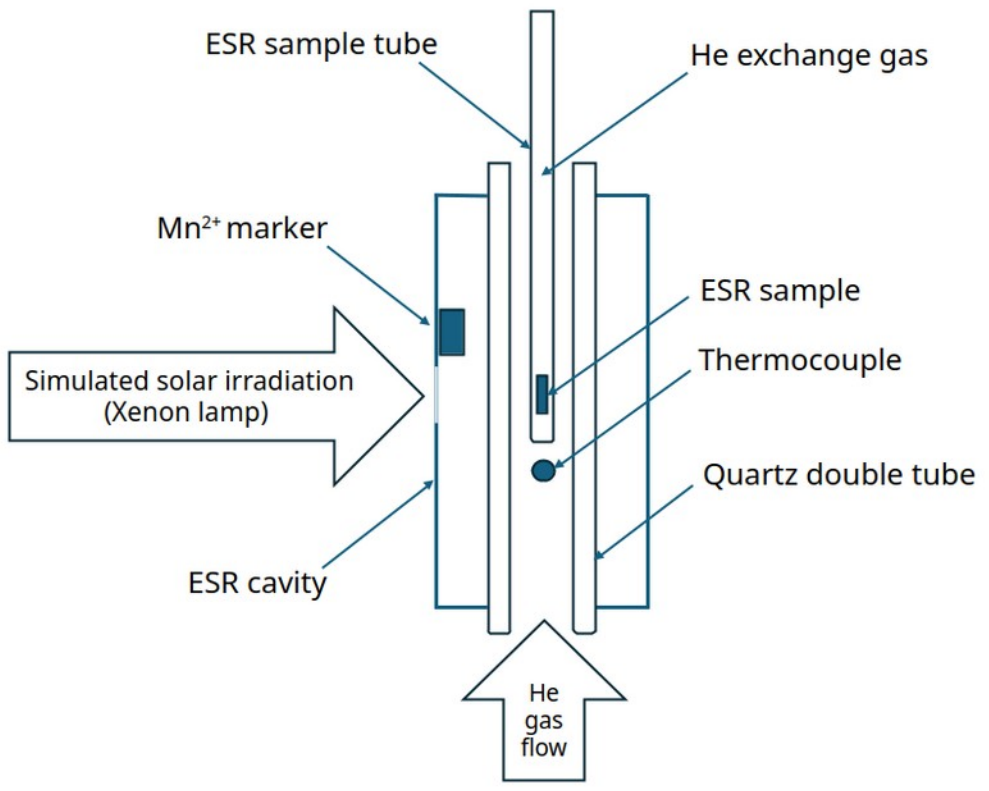

**Figure 1.** Scheme of the ESR measurement setup.

### 2.5. Theoretical Tools

Properties of the ground state such as the band structure, magnetic states and Bloch functions have been calculated with the density functional theory (DFT) [47,48] using the Quantum ESPRESSO package [49]. The generalized gradient approximation (GGA) has been chosen according to the Perdew-Burke-Ernzerhoff parameterization for the exchange-correlation functional [50]. Norm-conserving pseudopotentials have been used, with the energy cutoff for the plane waves set to 60 Ry. The first Brillouin zone (BZ) has been sampled using the uniform Monkhorst-Pack grid [51] with the *k*-point mesh of 12×12×4.

Excitonic properties have been obtained from the solution of the *ab initio* BSE (*ai*-BSE) [52] implemented in the *ab initio* many-body perturbation theory (*ai*-MBPT) code called Yambo [53,54]. The *ai*-BSE explicitly includes the electron-hole (e-h) correlations and takes the eigenvalues and eigenvectors of the DFT solution as building blocks of the perturbed Hamiltonian that describes the excited states. Tamm-Dancoff approximation has been used due to the computational complexity [55]. The e-h space contains 8 bands. The number of orbitals for the polarization function has been more than twice the number of electrons in the system. The plane waves have been used up to the energies of 10 Ry for the exchange components, 6 Ry for the screening and 4 Ry for response block size.

The bands interpolation has been done with the DFT Hamiltonian representation in the basis of the MLWF [56,57] using the wannier90 code [58]. Detailed analysis of the spin-bands involved in the optical transitions has been performed by careful examination of their composition with respect to the MLWF centered at the Cr atoms.

## 3. Results and Discussion

### 3.1. Characterization of the crystal structure and calculations of the ground state properties

The $Cr_2AlC$ MAX phase occurs in the hexagonal crystal structure with the space group $P6_3/mmc$ [59], and its $Cr_2C$ and Al layers are presented in Figure 2a for the side view and top view. Carbon atoms are 6-fold coordinated by three Cr atoms in the layers above and below the C layer, and this builds the $sp^3d^2$ hybridization around C involving electrons from the *3d* shell of Cr.

The DFT calculations lead to the metallic ground state with the energetically close magnetic phases: paramagnetic (PM), antiferromagnetic (AF) and ferromagnetic (FM), which are summarized in Table S1 in SI. Results depend on the parametrization of the exchange-correlation functional, pseudopotentials and used geometry with the experimental lattice parameters or these theoretically calculated. This means in practice that coexistence of phases at low temperatures is very likely. The AF order develops in $Cr_2C$ layers, with the local magnetic moment at each Cr of ~1 $\mu_B$ and the absolute value of the magnetic moment in the elementary cell, containing two $Cr_2C$ layers, is close to 5 $\mu_B$. Some partial spin moments are localized at Al atoms and their orientations are opposite to the magnetic moments at the adjacent Cr. Localization of the most pronounced positive and negative magnetic moments in the elementary cell is presented in Figure 2b that presents the 3D maps of the spin polarization in the AF phase.

Results of the $Cr_2AlC$ synthesis are shown in SEM images. Figures 2c and 2d present the powder mixture before and after heating, respectively, indicating a transformation of the material structure as a result of the thermal treatment. The SEM image of the product clearly confirms the layered structure that is characteristic for the MAX phase [59-62].

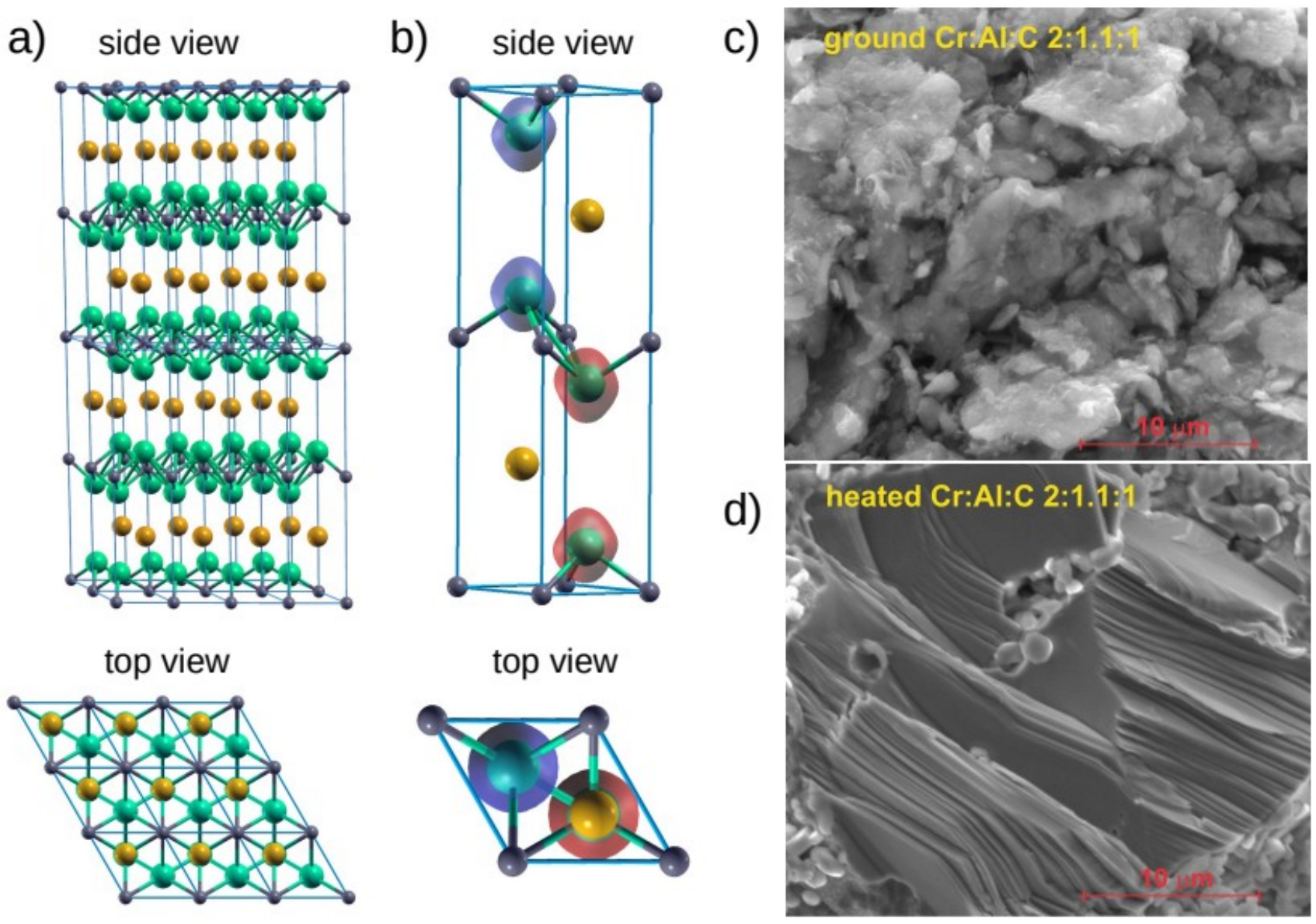

**Figure 2.** Crystal structure of $Cr_2AlC$. a) Atomic structure: side and top views for the periodic repetition of the elementary unit cell. Color legend: C – grey, Cr – green, Al – orange. b) Elementary cell with the 3D map of spin distribution, positive in red and negative in blue. c) SEM image of milled precursors mixture before heating. d) SEM image of synthesized $Cr_2AlC$ product after heating.

After the synthesis, the obtained material has been characterized. The x-ray diffractograms of the precursor mixture after milling, the synthesized product, and the $Cr_2AlC$ reference (PDF Card 290017) are shown in Figure 3a. Although carbon remains in the mixture after milling, no reflections characteristic of graphite have been observed. This is because grinding destroys its ordered structure leading to a formation of an amorphous phase. The diffractogram for the milled sample shows reflections corresponding to the *hkl* planes: (111), (200), (220), (311), (222), (400), (331), (420) for aluminium (PDF Card 040787) and (110), (200), (211), (220) and (310) for chromium (PDF Card 060694), respectively. The diffractograms show no additional reflections. This proves that grinding has no effect on the chemical composition of the mixture and no side processes occur during it. The diffraction pattern obtained for the synthesized $Cr_2AlC$ product is consistent with the PDF Card 290017 standard. Characteristic reflections for the $Cr_2AlC$ MAX phase have been observed at: 13.8, 27.9, 36.2, 36.9, 42.1, 42.2, 46.3, 56.9, 65.2, 77.1, 80.8, 111.0 and 114.8 2 Theta corresponding to the *hkl* lattice planes: (002), (004), (100), (101), (103), (006), (104), (106), (110), (109), (116), (209) and (213), respectively. This confirms that we obtain the MAX phase in the hexagonal crystal structure with the space group $P6_3/mmc$ [59,60]. The diffraction pattern shows additional, very low-intensity reflections marked with an asterisk, indicating the formation of a very small amount of impurities from other phases such as $Al_2O_3$ (nonmagnetic) or $Cr_7C_3$ (theoretically described as nonmagnetic [63]).

The high resolution TEM (HRTEM) images of the synthesised sample, presented in Figure 3b, demonstrate the presence of well organised areas of the crystal structure. The Fast Fourier Transform (FFT), presented in Figure 3c, and Inverse Fast Fourier Transform (IFFT), presented in Figure 3d, have been done for the red-framed regions of the HRTEM images, and both analyses correspond to the two crystal directions. These characterization techniques reveal that the distances between planes are of 0.641 nm and 0.248 nm corresponding to the (002) and (100), respectively, lattice planes of the hexagonal $Cr_2AlC$ [61,62]. The EDX study, presented in Figure 3e, definitively shows the presence of chromium, aluminium and carbon in the $Cr_2AlC$ sample. Additionally, the Cu peak comes from the copper mesh on which the sample is placed, while the small Si peak originates from the EDX Si-Li detector. In conclusion, structural and morphological studies confirm the efficiency of the synthesis of $Cr_2AlC$.

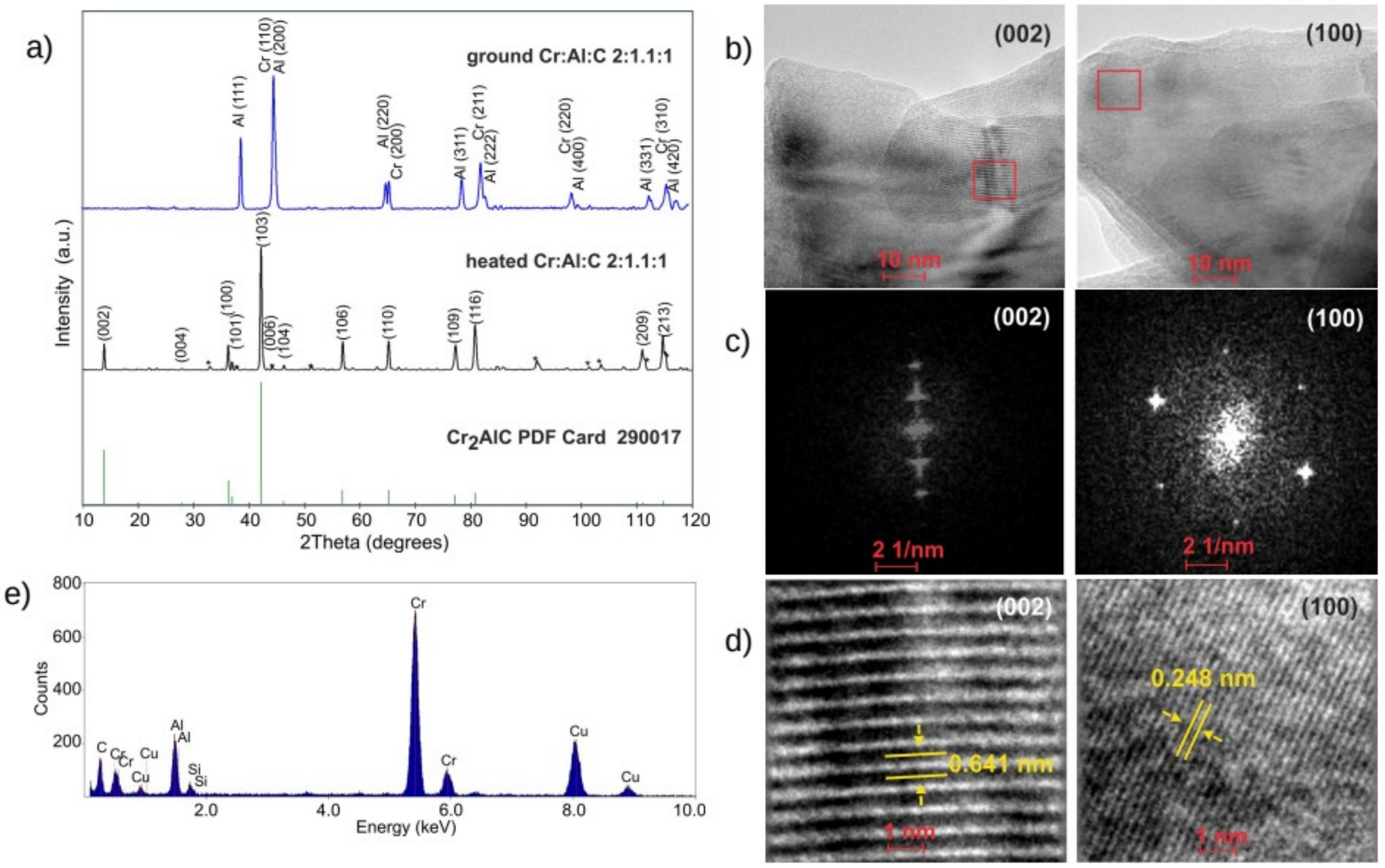

**Figure 3.** Characterization of $Cr_2AlC$. a) X-ray diffractograms of the milled precursors mixture (blue), synthesize product (black) and standard $Cr_2AlC$ PDF Card 290017 (green). b) HRTEM images of $Cr_2AlC$. c) FFT images obtained for the surface marked with a red frame in the corresponding HRTEM images. d) IFFT images obtained for the surface marked with a red frame in the corresponding HRTEM images. e) EDX profile for the synthesized $Cr_2AlC$.

### 3.2. Magnetic properties obtained with SQUID

The material exhibits clear paramagnetic properties in the whole studied temperature range as exemplified in Figure 4. At 300 K the magnetization $M$ is strictly linear in field within experimental accuracy, as presented in Figure 4a, with a mass susceptibility $\chi=3.5(1)\times10^{-6}$ emu/g/Oe. Since the diamagnetic component of $Cr_2AlC$ estimated from Pascal constants amounts to $\chi_{dia} \cong 5\times10^{-7}$ emu/g/Oe [64], we obtain the Cr-related susceptibility $\chi_{\mathrm{Cr}}$(300 K) of $4.0(2)\times10^{-6}$ emu/g/Oe. The corresponding magnetization at 10 kOe is $M_{Cr}$(300 K) = 0.040(2) emu/g, which gives $5.1(3) \times 10^{-4}$ $\mu_B$ per Cr atom, or $1.0(1) \times 10^{-3}$ $\mu_B$ per $Cr_2AlC$ formula unit.

Such a small, field-linear response is characteristic for Pauli paramagnetism of conduction electrons, $\chi_P$, and represents the dominant magnetic component in $Cr_2AlC$. This assignment is consistent with previous NMR and magnetometry studies that reported the Korringa-type relaxation and a weak temperature dependence of $\chi$ over a broad range of temperatures [65]. While Van Vleck paramagnetism can also yield a temperature-independent term, it typically arises from localized ionic states with low-lying excited levels. This scenario is unlikely in a good metal like $Cr_2AlC$ where hybridized Cr-*3d* states form itinerant bands at the Fermi level. Additionally, we note that no signatures of long-range magnetic order are detected what signifies a single magnetic constitution of the material synthesized for this studies.

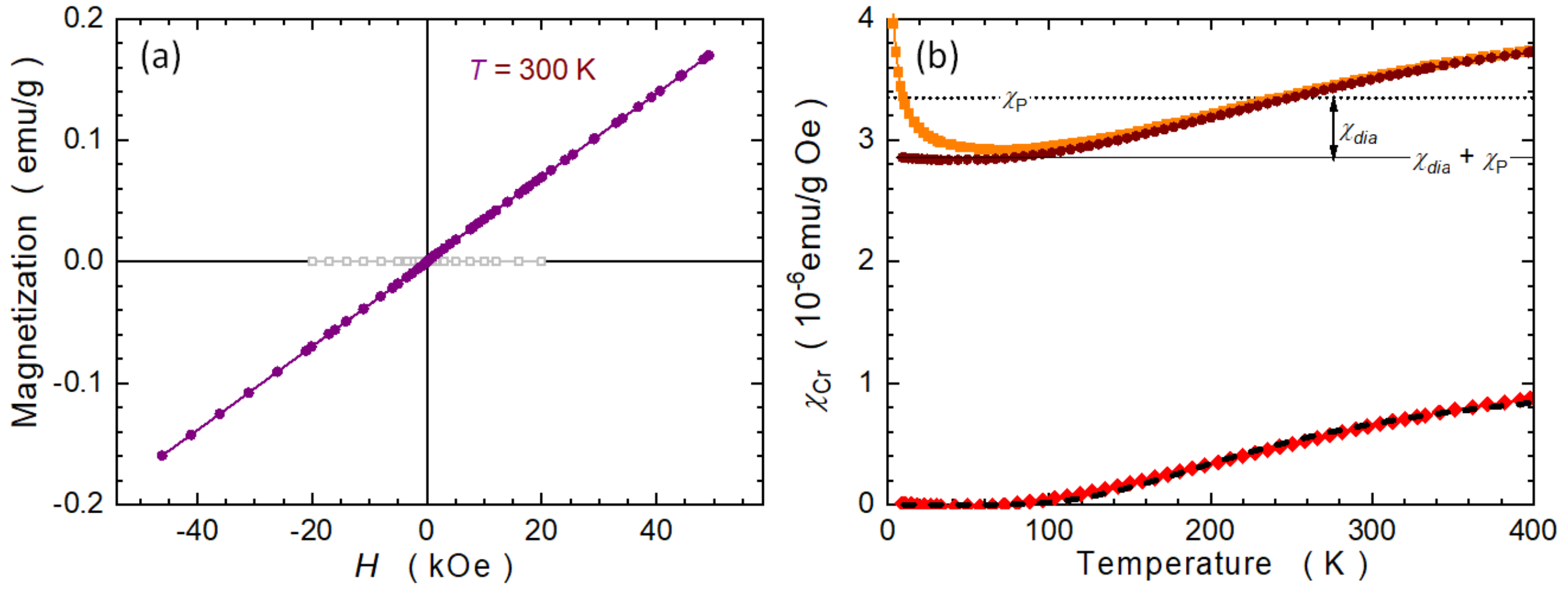

**Figure 4. a**) Magnetization of the $Cr_2AlC$ powder as a function of magnetic field *H*, measured at 300 K using the capsule compensating sample holder (CCSH) – full symbols – versus a test measurement of the empty capsule in the same CCSH – grey open symbols. **b**) Temperature dependence of the magnetic susceptibility $\chi_{Cr}$ of Cr subsystem in $Cr_2AlC$: Orange squares – as measured, brown full circles – after removing the Curie paramagnetism; details are given in SI, Sec. S3. The horizontal dotted line marks the magnitude of the Pauli paramagnetism after correction for the core diamagnetism of the constituent atoms, $\chi_{dia} \cong 5 \times 10^{-7}$ emu/g/Oe [66]. Full red diamonds mark the remaining *T*-dependent part of $\chi_{Cr}$, evaluated in the view of AF coupled Cr-Cr dimers and represented by the black thick dashed line.

On the other hand, as depicted in Figure 4b by orange squares, the measured magnetic susceptibility $\chi(T) = M(T)/H$ (with $H = 10$ kOe) reveals three distinct contributions which all together reproduce the overall temperature dependence. Listing in order of magnitude, these contributions are as follows: The first consists of two temperature-independent components discussed above, namely the Pauli paramagnetic susceptibility $\chi_P$ and the core diamagnetism of the constituent atoms $\chi_{dia}$. Their summation forms the temperature-independent baseline of $\chi(T)$, marked by the thin black line in Figure 4b. The second contribution is a Curie-type upturn that develops below 100 K. The third component is the term that increases with temperature and tends to saturate above 400 K.

The low-temperature paramagnetic component, $\chi_{PM}(T)$, is ubiquitously observed in MAX-phase compounds [67,68,34]. We quantify this contribution using differential Brillouin function analysis (DBFA) of magnetization curves measured between 2 and 8 K, as described in detail in SI, Sec. S3 (including Figs. S3 and S4, and references 66 and 69–72). The analysis reveals that the low-temperature upturn originates from a very small concentration, approximately 0.02%, of localized spins with $S = 3/2$. Using the spin quantum number and concentration determined by DBFA, we calculate the paramagnetic contribution $\chi_{PM}(T)$ over the entire experimental temperature range from the full Brillouin function evaluated at H = 10 kOe. This contribution is then subtracted from the measured susceptibility, allowing the remaining magnetic components to be analyzed indepen-dently. Importantly, this procedure does not require separate $M(H)$ measurements at every temperature, since the parameters describing the localized paramagnetic centers have already been established independently from the low-temperature magnetic isotherms.

We attribute these localized spins to a small fraction of Cr sites in structurally or chemically distinct environments, where electronic delocalization is locally reduced and local moments can be stabilized. A minute population of such centers may arise from antisite defects, off-stoichiometric regions, locally distorted Cr environments, surface-oxidized layers, or grain-surface terminations. In a powder sample, the latter contribution may be enhanced by the distribution of grain surfaces, terminations and defect configurations. The sensitivity of $Cr_2AlC$ to such local perturbations is also consistent with the near degeneracy of several magnetic/electronic configurations predicted by DFT using different parametrizations, summarized in Table S1 in SI, Sec. S2.

Determination of $\chi_{PM}$ allows us to isolate the temperature-independent contribution to the total susceptibility, $\chi_{P} + \chi_{dia}$, and thereby to extract the absolute value of the Pauli susceptibility $\chi_P \cong 3.3 \times 10^{-6}$ emu/g/Oe, as indicated by the dotted line in Figure 4b. Such a small $\chi_{P}$ is typical for MAX-phase compounds and other metallic systems, and enables even a very little concentration of localized paramagnetic spins to become discernible at low temperatures. In our sample, the $\chi_{PM}$ term contributes only by about 1 % to the total $\chi$ at 300 K, and thus has no significant influence on the room-temperature magnetic response. Therefore it can be neglected in the following discussion.

The remaining $T$-dependent component of $\chi_{Cr}(T)$, indicated by the solid red diamonds in Figure 4b, cannot be described by a simple Curie-Weiss or Pauli term. Notably, in the previous high-temperature study of $Cr_2AlC$ up to 800 K, a broad maximum of $\chi(T)$ centered around 350 - 400 K was observed [34]. In agreement with that report, our measurements also show the increase of $\chi(T)$ with temperature, although with a progressively reduced slope above 300 K suggesting that a similar maximum may occur in our sample at temperatures above 400 K, which is beyond our experimental limit.

Such a characteristic temperature dependence of the magnetic susceptibility can be described using the Bleaney-Bowers (B-B) model for antiferromagnetically coupled pairs of $S = 1/2$ spins. We adopt the isotropic exchange Hamiltonian $H = J_{eff}\, S_1 \cdot S_2$, where $J_{eff} > 0$ corresponds to antiferromagnetic coupling. Each spin pair has a singlet ground state ($S_{total} = 0$) and a thermally excited triplet state ($S_{total} = 1$), separated by an energy gap $\Delta = J_{eff}$. The residual temperature-dependent susceptibility is then fitted using:

$$\chi_{B-B}(T) = (C/T) \cdot \left[2/\left(3 + e^{J/k_B T}\right)\right] \quad \text{(Eq. 1)}$$

Here, $C = N_d \cdot (g \cdot \mu_B)^2 / k_B$ is the fitting prefactor, $N_d$ is the effective number of spin pairs per gram of sample, and $g = 2$ is assumed. The least-squares fit, represented by the thick dashed line in Fig. 4b, yields $J_{eff} / k_B = 760(10)$ K and $C = 1.60(2) \times 10^{-3}$ emu·K/g/Oe.

The fitted prefactor corresponds to $N_d = 6.4(2) \times 10^{20}$ spin pairs per gram. Within the independent-pair approximation, the corresponding fraction of Cr sites is $\xi = 2\, N_d / N_{Cr} = N_d / N_{f.u.} = 0.15(1)$, where $N_{Cr} = 2\, N_{f.u.}$ is the number of Cr atoms per gram of $Cr_2AlC$. This fraction represents the effective population of antiferromagnetically coupled Cr-Cr spin pairs within the B-B model. The analysis thus provides a quantitative description of the additional susceptibility in terms of local AFM pair correlations, although their spatial arrangement and possible dynamics cannot be established from the present magnetometry measurements.

The assumed spin value $S = 1/2$ is consistent with the calculated local Cr spin polarization of approximately 1 $\mu_B$ in the AFM configuration. However, this correspondence should be regarded as approximate, given the itinerant character of the Cr-derived electronic states. This value may be slightly underestimated by the DFT owing to its approximate treatment of the electron correlations and a way of partitioning a crystal to atoms. Formal oxidation state of Cr in $Cr_2AlC$ is not uniquely defined, a one-to-one correspondence between an ionic configuration and the observed spin state cannot be established [75]. Thus, the main magnetic response is ascribed to a weak metallic Pauli-like susceptibility modified by substantial local AFM pair correlations described by the B–B model. By contrast, the dilute Curie-like component represents only a minute fraction of Cr sites and is relevant mainly for the low-temperature upturn.

For comparison, we report both the conventional sample-averaged electronic density of states and a conditional estimate normalized by $f_{active}$ = 1-$\xi$ = 0.85. The latter assumes that the Cr fraction represented by the effective B-B contribution can be excluded from the itinerant subsystem; it should therefore be regarded as a model-dependent normalization rather than an independently established separation of Cr sites.

Table 1. Material parameters derived from the Pauli susceptibility differentiating two different approaches for the whole sample and conditional active-fraction normalization (assuming active-fraction $f_{active}$ = 1- $\xi$ = 0.85). The values given for the effective density of states (EDOS) at the Fermi level, EDOS($E_F$) = $\chi_P/\mu_B^2$, indicate their upper limits in $Cr_2AlC$.

| Quantity | Symbol (unit) | Sample-averaged ($f_{active}$ = 1.00) | Active-fraction normalized ($f_{active}$ = 0.85) |
|---|---|---|---|
| Molar susceptibility | $\chi_m$ ($cm^3$/mol) | 4.7(1)×$10^{-4}$ | 5.5(1)×$10^{-4}$ |
| Effective density of states for both spins | EDOS($E_F$) (states·$eV^{-1}$·f.u.$^{-1}$) | 15(1) | 17(1) |
| Effective density of states per Cr atom | EDOS($E_F$)/Cr (states·$eV^{-1}$·$Cr^{-1}$) | 7(1) | 10(1) |

An illumination-dependent SQUID measurement was performed as a wavelength-specific optomagnetic test using a red LED (λ ≈ 610 nm, approx. 2.03 eV). Within the experimental sensitivity, no intrinsic change in the macroscopic magnetic moment was detected. The only reproducible variation was consistent with a small illumination-induced temperature offset, as discussed in SI, Sec. S1. Importantly, this negative result applies specifically to the investigated illumination conditions and does not exclude a magnetic response at other excitation wavelengths.

### 3.3. Magnetic properties and optical switching obtained with ESR measurements and BSE calculations

ESR measurements are first performed on a powder sample of $Cr_2AlC$ at room temperature. No ESR signal has been detected either in the dark or under light irradiation. To probe possible low-temperature magnetic centers, ESR measurements are then carried out at 4 K. As shown in Figure 5a, a clear ESR signal with a $g$-factor of 2.0038 is observed in the dark state. Calibration against the $Mn^{2+}$ standard gives $N_{spin}$ = 3.80 × $10^{17}$ spins per gram of $Cr_2AlC$, which corresponds to only 0.0045% of all Cr sites. The peak-to-peak linewidth, $\Delta H_{pp}$, is 0.92 mT. The ESR signals of the $Mn^{2+}$ standard appear with the opposite phase to those of $Cr_2AlC$ because the microwave oscillating magnetic fields at the positions of the standard and the sample have opposite directions.

After 1 h of simulated solar irradiation, the ESR signal intensity decreases, the $g$-factor changes from 2.0038 to 2.0051, and $N_{spin}$ decreases to 8.58 × $10^{16}$ spins per gram. At the same time, $\Delta H_{pp}$ increases to 1.32 mT. After switching off the irradiation, the signal intensity, $g$-factor and linewidth return to their initial dark-state values, as shown in Figure 5b. The optically modified part of the ESR signal therefore corresponds to about 2.90 × $10^{17}$ spins per gram, i.e. only ~0.0035% of all Cr sites. Thus, ESR reveals a reversible light-induced modification of a very dilute population of local magnetic centers in $Cr_2AlC$, rather than a bulk optical switching of magnetism in the MAX phase.

A plausible microscopic mechanism for the reduction of the local ESR signal is light-induced redistribution of spin polarization between adjacent Cr sites. This scenario is schematically shown in Figure 5c. In regions where neighboring Cr atoms carry opposite local spin polarizations, optical excitation may transfer electronic weight from a majority-spin state localized on one Cr site to states with larger contribution from the neighboring Cr site with the opposite local moment. In this simplified picture, optical excitations from the ground state (G) to excited states (E) in a given spin channel modify the spatial distribution of spin polarization through the different atomic composition of the involved Bloch states, as illustrated in Figure 5d by the $G_{ij}$ and $E_{ij}$ components.

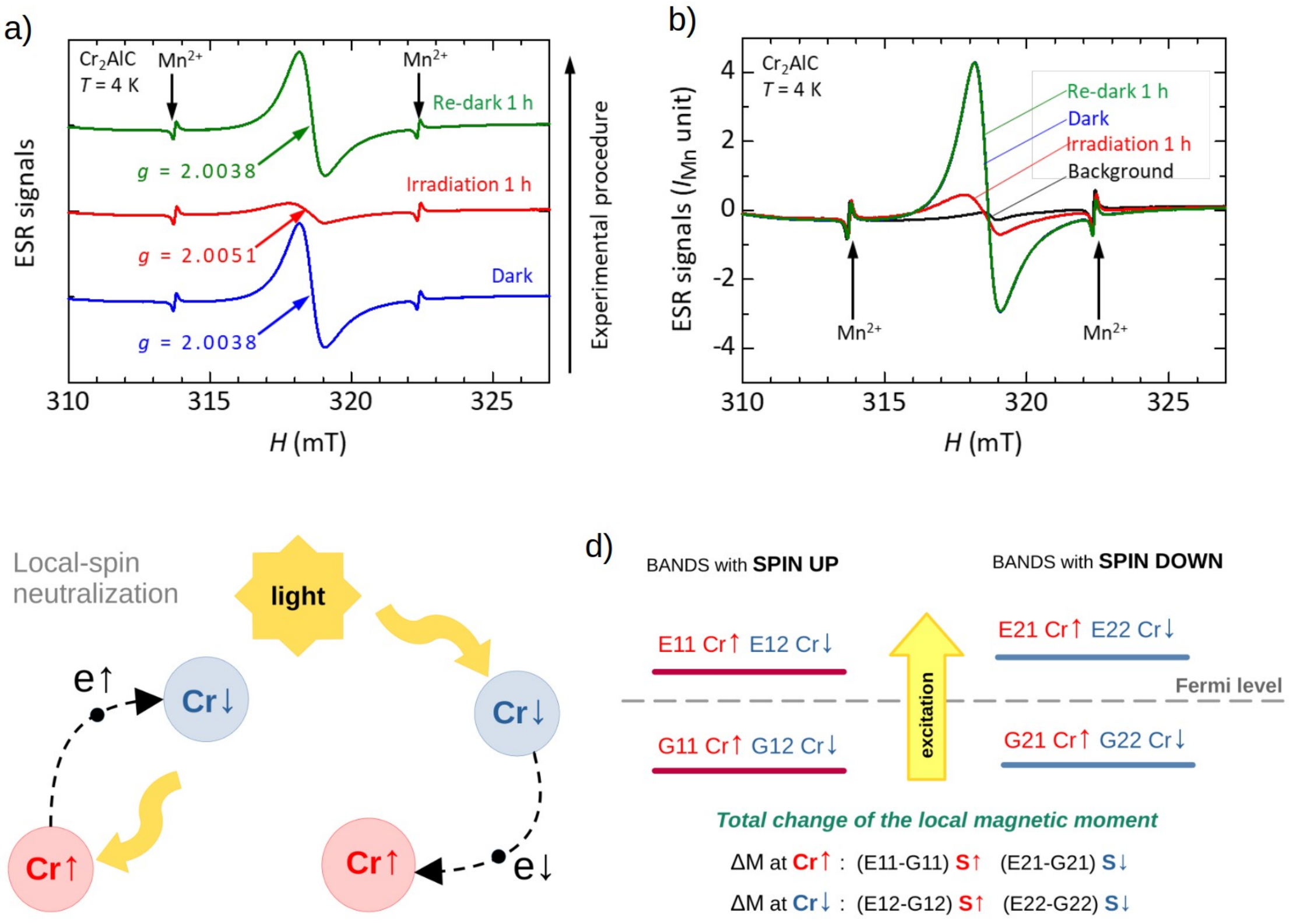

**Figure 5.** Change of the local magnetization due to light excitation. a) ESR signals of $Cr_2AlC$ at 4 K with experimental procedure: in the dark state – blue line, under light irradiated for 1 h – red line, in the dark state for 1 h after the irradiation – green line. b) Comparison of ESR signals presented in (a) and the background signal of the ESR cavity resonator – black line. In (a) and (b), ESR signals from the standard $Mn^{2+}$ sample are shown. c) Scheme of the spin hopping, denoted by e↑ and e↓, between the Cr atoms with the opposite local moments, denoted by Cr↑ and Cr↓, induced by light absorption. d) Scheme of the MLWF-composition analysis of the spin-bands involved in the optical transitions. States below and above the Fermi level are denoted by G and E, respectively, first and second index are denoted spin and atomic-shell, respectively, and the change of local magnetization is denoted by ΔM.

We are mostly interested in bands composition with respect to the Cr atoms, therefore we present the projections of the spin-bands up and down on the MLWF centered at two Cr with the local magnetic moments up and down in Figure 6. Definitions of the high symmetry points in the Brillouin zone (BZ) sampling for the hexagonal crystal structure are given in Table S2 in SI. For completeness, projections of the spin bands on the MLWF centered at Al and C are presented in Figure S5 in SI. Electrons from the atomic shells of Al and C do not contribute to the band structure in the energy region of ±2 eV below and above the Fermi level, thus being optically inactive.

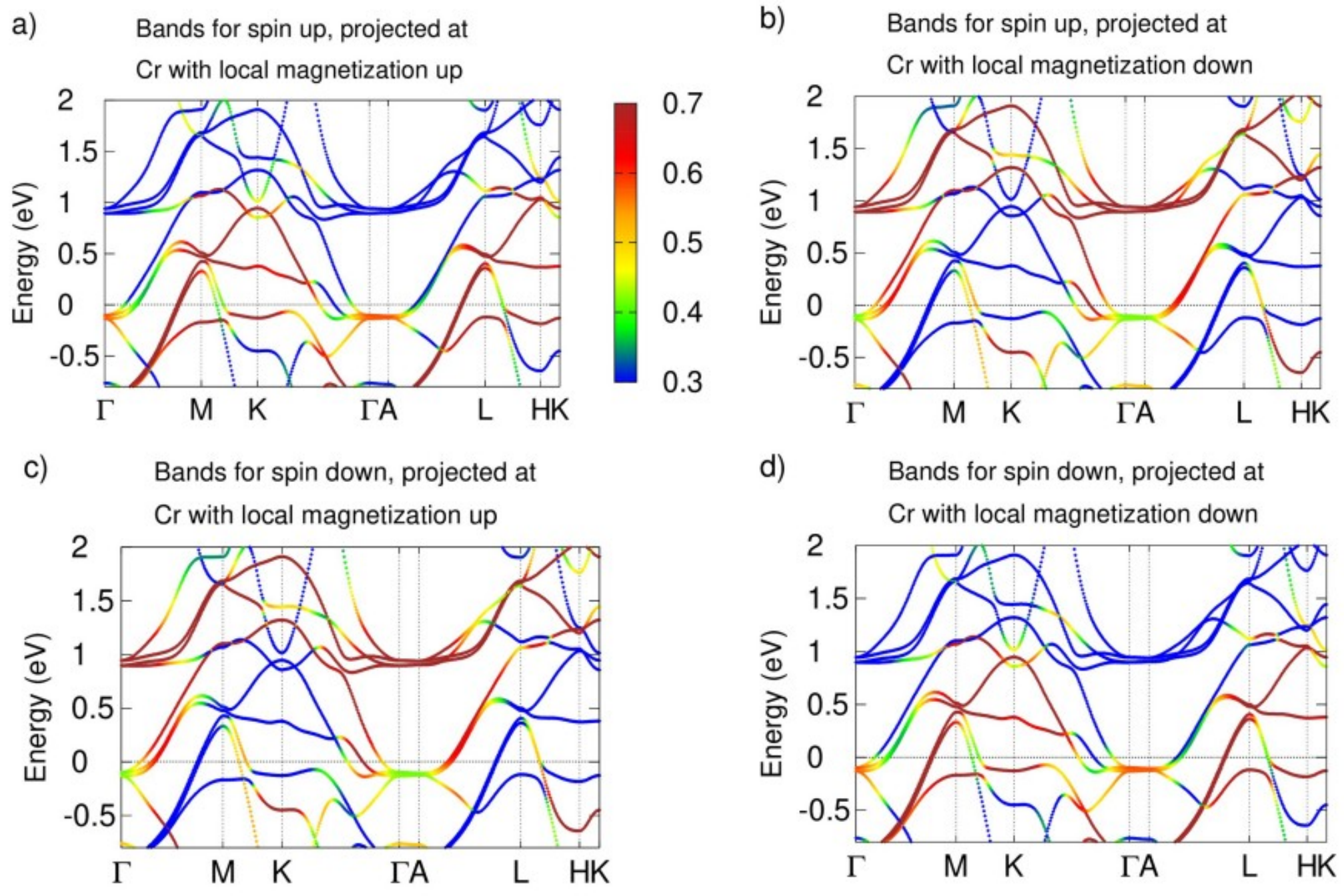

**Figure 6.** Spin-bands projected at the MLWF centered at two Cr atoms with the local magnetization up and down. Color legend is common for all cases and truncated to the range [0.3,0.7] for a better contrast. a) Bands for spin up projected at Cr with local magnetic moment up, Cr↑. b) As in (a) for Cr with the local magnetic moment down, Cr↓. c) and d) correspond to a) and b), respectively, for bands with spin down.

Examining closely the Γ-A high symmetry line in BZ in Figures 6a-d: the majority spin-bands at each type of Cr (*i.e.* spin up for Cr↑ and spin down for Cr↓) are localized below the Fermi level, while the minority spin-bands (*i.e.* spin down for Cr↑ and spin up for Cr↓) are localized above the Fermi level. Similar situation is partially for M-Γ and L-A high symmetry lines. Thus, the direct optical excitations in these regions of the BZ would reduce the local magnetic moment at each Cr. In contrast, the opposite composition of the spin-bands is near the M point towards K and near the L point towards H, therefore the optical transitions in these regions would lower the local magnetic moments at Cr, while transitions near K would be neutral for a change of the local magnetization.

Each exciton consists of several nearly degenerate dipole transitions, *i.e.* band-to-band excitations at a given *k*-point in BZ. Composition of bands involved in the optical transitions with respect to the atomic shells of four Cr atoms in the elementary cell indicates an effect of light absorption of given energy for a change of the local magnetization, which we define as

$$\Delta M = \Sigma_{Cr\uparrow,Cr\downarrow}\ \Sigma_{v,c}\ \Sigma_{\mathbf{k}}\ \omega_{\mathbf{k}} \{[A_{c\uparrow,\mathbf{k}}(Cr\uparrow) - A_{v\uparrow,\mathbf{k}}(Cr\uparrow)] - [A_{c\downarrow,\mathbf{k}}(Cr\uparrow) - A_{v\downarrow,\mathbf{k}}(Cr\uparrow)] + [A_{c\downarrow,\mathbf{k}}(Cr\downarrow) - A_{v\downarrow,\mathbf{k}}(Cr\downarrow)] - [A_{c\uparrow,\mathbf{k}}(Cr\downarrow) - A_{v\uparrow,\mathbf{k}}(Cr\downarrow)]\} \quad (2)$$

The coefficients $A_c$ and $A_v$ denote contributions of the MLWF centered at Cr to the conduction band, *c*, and valence band, *v*, and arrows in the subscript indicate the spin channel of the corresponding band while arrows near Cr indicate the local magnetic moment at a given atom. We sum this formula over all atoms with the same local magnetic moment, valence and conduction bands, and ***k***-points with their weights $\omega_{\mathbf{k}}$ in BZ.

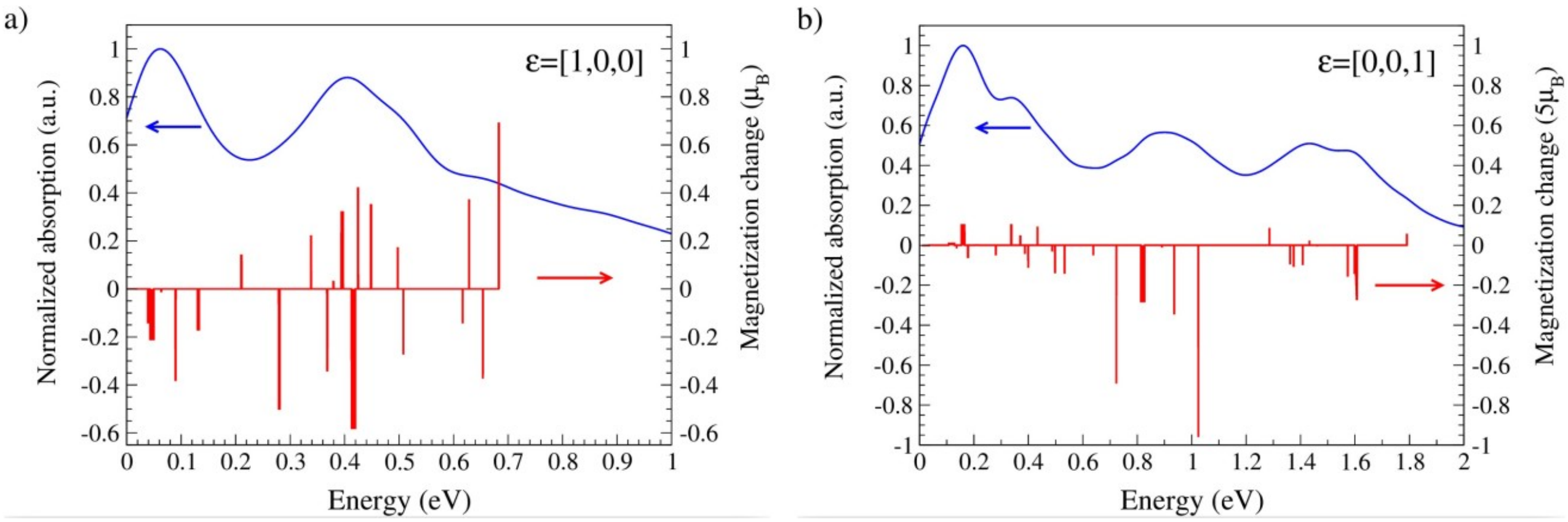

**Figure 7.** The normalized absorption spectrum obtained from the *ai*-BSE (blue line) and change of the local magnetization at all Cr atoms induced by an exciton of given energy (red bars) for the electric-field vectors: a) [1,0,0] and b) [0,0,1]. Color arrows point to the proper axis at each plot. Note that the scale of the magnetization axis is 5 larger for [0,0,1] than for [1,0,0].

The normalized absorption spectra obtained from the *ai*-BSE for two light polarizations are presented in Figure 7a-b, that contains also bars corresponding to the summed change of the local magnetization at all Cr atoms induced by excitons of a given energy. Thickness of these bars indicates the nearly degenerate optical transitions. Please note that the change of magnetization in a case of the light polarization [0,0,1], in Figure 7b, is plotted in units that are five times larger than those for the polarization [1,0,0], in Figure 7a. Effectiveness of the optical switching of local magnetic moment at a given wavelength depends on two factors: *i)* the number of nearly degenerate excitons and their oscillator strengths and *ii)* the favorite compositions, with respect to the Cr atomic shells, of the spin bands involved in the dipole transitions. In order to give a full picture of **the optical switching effects,** we include additional data in SI: 1) the absolute (not normalized) absorption spectra for the [1,0,0] and [0,0,1] polarizations are presented in Figure S6, 2) the list of *k*-points used in the irreducible wedge of BZ is in Table S3, and 3) the list of excitons with the oscillator strengths larger than 0.1 is given in Table S4 including the order number of given exciton in the spectrum and composition of the main dipole transitions found in the corresponding eigenvectors (it means the order numbers of involved bands, their spin channel and *k*-point), as well as the change of the local magnetic moment at all Cr in the cell induced by the proper dipole transition, 4) the comparative plot of light-induced local magnetic moments with respect to the light energy and the exciton oscillator strengths, in Figure S7, for both light polarizations: a) [1,0,0] and b) [0,0,1].

Taking into account all data, we summarize, that the most effective parts of light absorption for switching the magnetic moment are the strongest in the infrared part of the spectrum (near 1200 nm) for the light polarization along the crystal *c*-axis (*i.e.* crystal direction [0,0,1]) due to the reversed spin-band compositions with respect to the majority spin at the Cr atoms. Some less effective switching could occur in the THz region of the spectrum with the $Cr_2C$ in-plane polarization (*i.e.* crystal direction [1,0,0] and the symmetry equivalent [0,1,0]) due to high oscillator strengths and magnetization-reducing effect. However, the present *ai*-BSE calculations describe interband optical excitations and do not explicitly include the intraband Drude response expected in metallic $Cr_2AlC$. Consequently, the predicted low-energy spectral features should not be interpreted as a complete description of the THz optical absorption. We emphasize the calculated spin-redistribution response associated with the interband excitations, particularly in the near-infrared range.

The different responses observed in the ESR and SQUID experiments are not necessarily contradictory. ESR probes the resonance response of a very dilute population of localized magnetic centers, whereas SQUID magnetometry measures the total magnetic moment of the sample. Moreover, the two experiments were performed under different illumination conditions. The ESR measurements employed broadband simulated solar irradiation (AM1.5G), while the SQUID optomagnetic test was restricted to a red LED emitting at approximately 610 nm (2.03 eV). Our theoretical calculations indicate a particularly pronounced local spin-redistribution response near 1 eV for light polarized along the crystallographic c-axis, a spectral region not directly accessed by the red LED. Consequently, the absence of a detectable SQUID response under 610 nm illumination does not contradict the reversible light-induced modification of the ESR signal. However, the present experiments do not establish which spectral components are responsible for the ESR response.

To the best of our knowledge, illumination-dependent magnetic measurements have not previously been reported for $Cr_2AlC$. Nevertheless, optically induced intersite spin transfer has been theoretically demonstrated in other magnetic materials, including antiferromagnetically coupled systems, using time-dependent first-principles calculations [76]. Furthermore, wavelength-dependent light-induced ESR responses have been experimentally observed in $Cr_2O_3$ under laser illumination, demonstrating the importance of excitation wavelength in probing local magnetic responses [77]. These studies provide useful context for our broadband ESR measurements and the wavelength-specific SQUID optomagnetic test.

## 4. Conclusion

In this work, we synthesized the $Cr_2AlC$ MAX phase by annealing pellets prepared from elemental powders for 15 h at 1450 °C under an Ar pressure of 100 MPa, using heating and cooling rates of 600 °C/h. The obtained material was characterized by x-ray diffraction, energy-dispersive x-ray spectroscopy, scanning electron microscopy, and high-resolution transmission electron microscopy.

Magnetometry identifies $Cr_2AlC$ as a weak, field-linear paramagnet over the investigated temperature range, with the dominant response assigned to the Pauli-like susceptibility of itinerant Cr-derived states. The temperature dependence is, however, not purely metallic. After subtracting the small low-temperature Curie-like contribution, the remaining non-monotonic susceptibility is well described by the Bleaney–Bowers model for antiferromagnetically coupled $S = 1/2$ spin pairs. Within this model, the fitted amplitude corresponds to an effective population involving approximately 15% of the Cr sublattice, consistent with a substantial contribution from local AFM correlations.

The wavelength-specific SQUID optomagnetic test performed under 610 nm illumination revealed no detectable intrinsic change in the macroscopic magnetic moment within the experimental sensitivity. The only observed response was consistent with minor illumination-induced heating. These results are complementary to ESR measurements and DFT and *ai*-BSE calculations. ESR measurements at 4 K reveal a local magnetic signal that is strongly reduced under illumination and recovers in the dark. Its mechanism can be ascribed by the *ai*-BSE solution and the MLWF analysis as a spin excitation from Cr to the neighboring Cr atom. Importantly, the optically modified ESR signal corresponds only to several tens of ppm of the Cr sublattice. Thus, ESR detects a highly light-sensitive but extremely dilute population of local magnetic centers, whereas the overwhelming majority of Cr sites does not contribute to the detected optical switching. This naturally explains why the corresponding effect is not resolved as a macroscopic magnetization change in the magnetometry.

DFT calculations point to a metallic state with closely competing magnetic configurations, including paramagnetic and antiferromagnetic solutions. Taken together, the results suggest that $Cr_2AlC$ is not a conventional long-range ordered magnet, but a metallic system close to local magnetic instabilities. In this picture, itinerant Cr-derived states coexist with local antiferromagnetic Cr-Cr correlations and a trace concentration of localized paramagnetic centers. The optical sensitivity observed by ESR should therefore be understood as a local magnetic/electronic response of these rare centers or domains rather than as a bulk optical switching of magnetism in the $Cr_2AlC$ phase.

**Author contributions**

Małgorzata Wierzbowska: calculations and writing, Anna Basa: characterization and writing, Kazuhiro Marumoto: ESR and writing, Jiaxi Wang: ESR, Katarzyna Gas and Maciej Sawicki: SQUID and writing, Kacper Sierakowski: synthesis, Karel Carva: conceptualization (theory) and writing, Michał Boćkowski: synthesis and writing

**Conflicts of interest**

There are no conflicts to declare.

**Acknowledgements**

We gratefully acknowledge Poland's high-performance infrastructure PL-Grid in the Cyfronet Center for providing computer facilities and support within the computational grants No. PLG/2022/015264 and PLG/2022/016016. K.C. acknowledges Czech Science Foundation grant no. 23-04746S and Project Quantum materials for applications in sustainable technologies (QM4ST), funded as project no. CZ.02.01.01/00/22_008/0004572 by P JAK, call Excellent Research by Ministry of Education, Youth and Sports of the Czech Republic.

**Supporting Information**
Supporting Information is available from the Royal Society of Chemistry Online Library or from the author.

## References in Supporting Information

**Supporting information**

# Local magnetic correlations and light-sensitive centers in the $Cr_2AlC$ MAX phase

Małgorzata Wierzbowska[1,*], Anna Basa[2], Kazuhiro Marumoto[3,4,5], Jiaxi Wang[3], Katarzyna Gas[6,7,8], Maciej Sawicki[7], Kacper Sierakowski[1], Karel Carva[9], Michał Boćkowski[1]

[1] Institute of High Pressure Physics Polish Academy of Sciences, Sokolowska 29/37, 01-142 Warsaw, Poland
[2] Department of Chemistry, University of Bialystok, Ciolkowskiego 1K, 15-245 Bialystok
[3]Department of Materials Science, Institute of Pure and Applied Sciences, University of Tsukuba, Tsukuba, Ibaraki 305-8573, Japan
[4] Research Center for Organic-Inorganic Quantum Spin Science and Technology (OIQSST), University of Tsukuba, Tsukuba, Ibaraki 305-8573, Japan
[5] Tsukuba Research Center for Energy Materials Science (TREMS), University of Tsukuba, Ibaraki 305-8571, Japan
[6] Center for Science and Innovation in Spintronics, Tohoku University, 2-1-1 Katahira, Aoba-ku, Sendai 980-8577, Japan
[7] Institute of Physics, Polish Academy of Sciences, Aleja Lotnikow 32/46, PL-02668 Warsaw, Poland
[8] Laboratory for Nanoelectronics and Spintronics, Research Institute of Electrical Communication, Tohoku University, 2-1-1 Katahira, Aoba-ku, Sendai 980-8577, Japan
[9] Charles University, Faculty of Mathematics and Physics, DCMP, Ke Karlovu 5, 12116 Prague 2, Czech Republic

[*] corresponding author – Małgorzata Wierzbowska
e-mail: wierzbowska@unipress.waw.pl, Phone: +48 22 6325010, Fax: +48 22 6324218

## S1. Illumination setup for SQUID magnetometry

To enable magnetization measurements under illumination, an alternative sample-mounting approach was used. This configuration also allowed us to extend the measurement temperature range up to 400 K, since it does not involve thermoplastic straws, which soften above approximately 340 K. Several milligrams of $Cr_2AlC$ powder were deposited onto a silicon platelet with dimensions of approximately $4 \times 5 \times 0.2$ $mm^3$ and immobilized using highly diluted GE varnish [1], which has a negligible magnetic contribution. The complete specimen, consisting of the Si platelet with attached $Cr_2AlC$ powder, was then fixed with the same GE varnish to a silicon stick, approximately 8 inches long and 2 mm wide, which is the standard sample holder used in our MPMS setup [2]. This mounting configuration provides good thermal and mechanical stability [3-5], while allowing direct optical access to the powder surface. During the optimization of the SQUID measurements, several preliminary tests have been performed using different portions of the same powder. Importantly, the separate portion used for illumination measurements has reproduced, during cooling, the characteristic temperature dependence of the magnetic response previously observed for the much larger capsule-mounted specimen.

For illumination-dependent measurements, a red LED was mounted at the upper end of the Si-stick holder, approximately 7 cm above the $Cr_2AlC$ powder, as shown in Fig. S1. The emission wavelength of the diode was about 610 nm at 4 K. The LED was powered through two thin Cu wires connected to an external current source. The wires were routed to the laboratory environment through a custom-modified vacuum-tight electrical feedthrough integrated into the RSO head of the magnetometer.

To minimize the risk of insulation damage during reciprocating motion in the RSO mode, in particular possible shearing or abrasion of the wire enamel, the MaxSlope variant of the RSO mode was used for illuminated measurements [6]. In this mode, the sample travel distance is reduced from the standard 4 cm to 0.5 cm. A comparative test confirmed that magnetic moment values obtained in the standard full-scan RSO mode and in the MaxSlope mode agree within the experimental uncertainty.

The LED was typically operated at a current of approximately 1.5 mA. This was the highest current used in the experiment and was selected as a compromise between sufficient optical intensity at cryogenic temperatures and limited heating of the sample environment. Higher LED currents have not been investigated because the illumination-induced heating, already detectable under the applied conditions (Fig. S2), would increasingly compromise the reliability of the magnetization measurements.

The final illumination test, which also provides an estimate of the magnitude of LED-induced heating, is shown in Fig. S2. The illumination setup was designed to perform a wavelength-specific optomagnetic test under cryogenic conditions, rather than a wavelength-resolved optical excitation study. The choice of the red LED was dictated by the practical constraints of integrating an optical source into the SQUID magnetometer, particularly the limited available space and the need to maintain the cryogenic environment. The resulting experiment probes the magnetic response under illumination at approximately 610 nm (2.03 eV). This approach provides a direct experimental assessment of the macroscopic magnetic response to optical excitation under well-defined cryogenic conditions.

The LED-on trace, measured on cooling, is shifted with respect to the LED-off reference measured on warming. Although this appears as a downward shift of the magnetic moment, the effect is more naturally interpreted as a small shift of the illuminated data toward lower nominal temperatures. Under illumination, the sample requires additional cooling power from the environment to compensate for the heat generated by the incoming radiation. Owing to the steep slope of m(T) in the liquid-helium temperature range, caused by the Curie-like contribution (part S3), the $Cr_2AlC$ sample effectively acts as a sensitive thermometer. The observed displacement can be reproduced by a small temperature offset of about 0.3 K at 2 K and less than 0.05 K at 5 K.

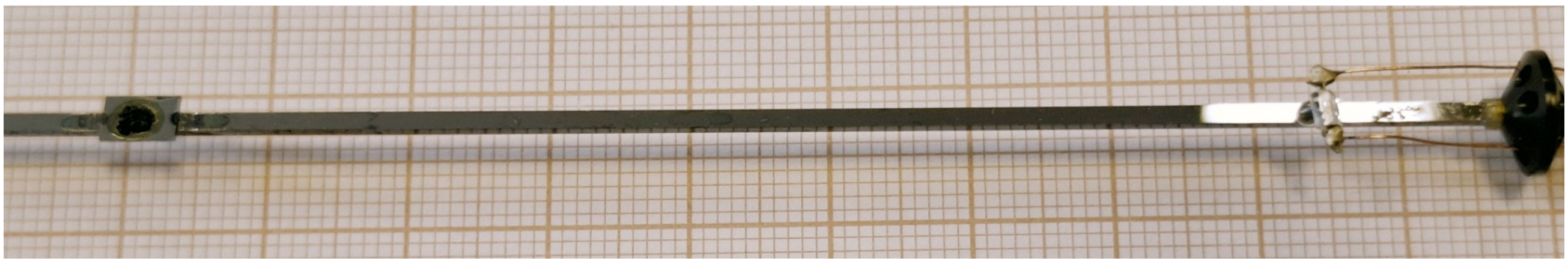

**Figure S1.** Photograph of the illumination-compatible SQUID magnetometry setup. The Cr2AlC powder was immobilized on a Si platelet attached to the Si-stick sample holder. A red LED was mounted above the sample and connected by thin Cu wires routed to the custom vacuum-tight feedthrough integrated into the magnetometer RSO-transport head.

This temperature offset remained sufficiently small for testing possible optomagnetic effects. However, within the experimental sensitivity, no additional light-induced magnetic contribution was detected; the only observable outcome of illumination was the small, reversible, heating effect described above.

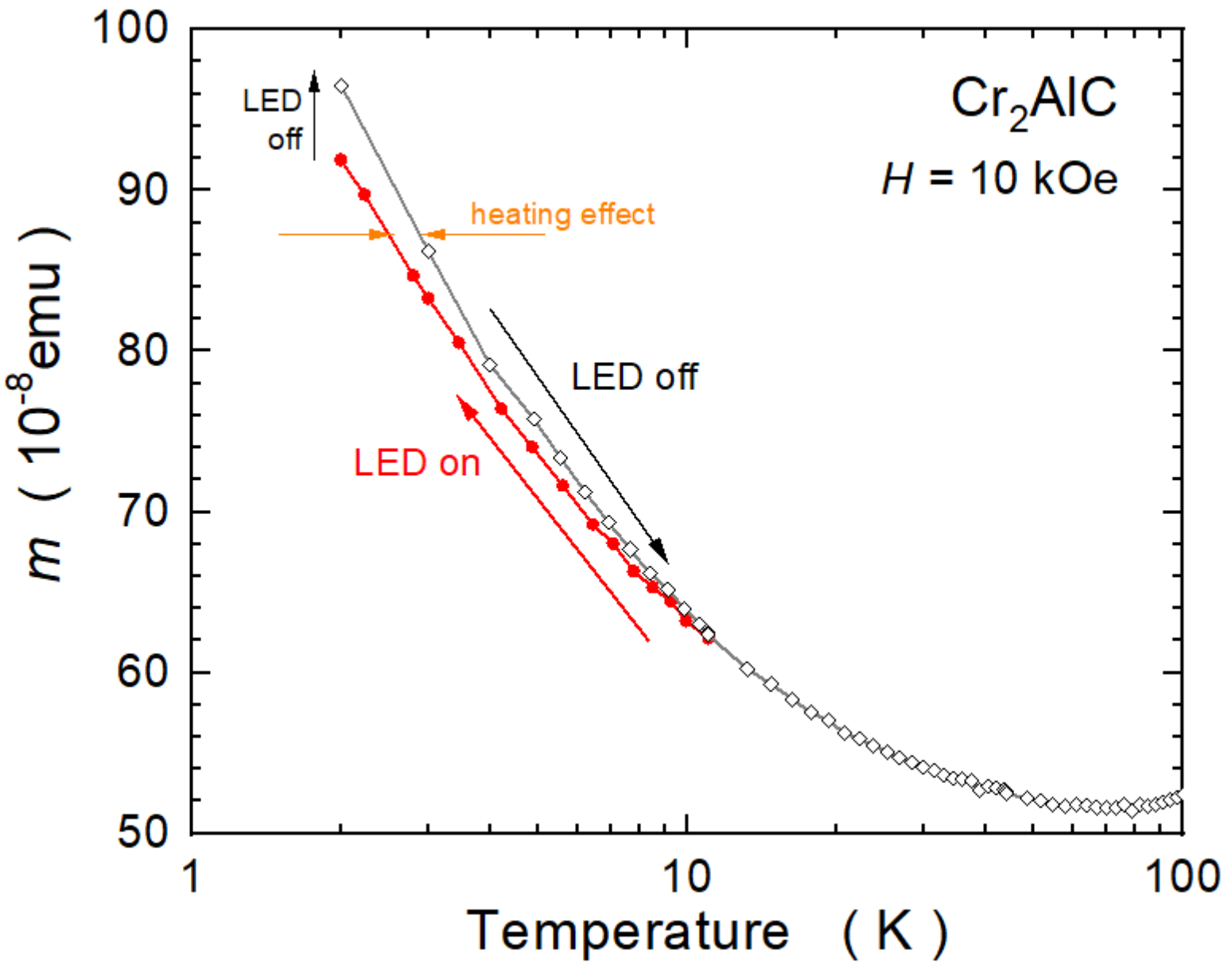

**Figure S2**. Temperature dependence of the magnetic moment of $Cr_2AlC$ measured at $H$ = 10 kOe with the red LED switched on and off. The small displacement of the LED-on trace with respect to the LED-off reference is largest at low temperature, where dm/dT is steepest, and is consistent with minor illumination-induced heating. Within the experimental sensitivity, the data do not indicate an intrinsic light-induced magnetic contribution beyond this heating effect.

**S2. Magnetic phases obtained from the DFT and their relative total energies**

Definitions of magnetic moments via the spin polarized electronic densities ρ↑(**r**) and ρ↓(**r**).

$M_{tot}$ – total magnetic moment – ∫ ρ↑(**r**) d**r** – ∫ ρ↓(**r**) d**r**.

$M_{abs}$ – absolute magnetic moment – is defined as: ∫ |ρ↑(**r**) – ρ↓(**r**)| d**r**.

**Table S1.** Magnetic phases of $Cr_2AlC$. Total energies, ΔE in meV, with respect to the paramagnetic state, total and absolute magnetic moments in in the cell (per 4 Cr atoms), $M_{tot}$and $M_{abs}$ in $\mu_B$, and relaxed lattice parameters, *a* and *c* in Å, obtained from the DFT with and without the spin-orbit coupling (SOC) employing the Perdew-Burke-Ernzerhof (PBE) or Perdew-Zunger (PZ) exchange-correlation functionals. Abreviations: PM – paramagnetic, AF – antiferromagnetic, FM – ferromagnetic phase. Experimental lattice vectors are *a* = 2.910 Å and *c* = 13.097 Å.

| $\Delta E=E_X-E_{PM}$, X=AF, FM ($M_{tot}/M_{abs}$) | | |
|---|---|---|
| geometry | experimental | relaxed lattice parameters |
| PBE, noSOC, Cr PP with 6 valence electrons | | |
| AF | -107 (0.0/5.06) | -64 (0.0/4.98), *a*=2.912, *c*=12.927 |
| FM | -117 (3.40/4.23) | -78 (3.37/4.15), *a*=2.897, *c*=12.950 |
| PZ, both noSOC and SOC, Cr PP with 6 valence electrons | | |
| AF/FM | Converges to PM | |
| PBE, noSOC, Cr PP with 6 valence electrons, Hubbard $\mathbf{U_{eff}=1\ eV}$ at Cr-3d | | |
| AF | -878 (0.0/10.11) | -1155 (0.0/11.08), *a*=3.024, *c*=13.061 |
| FM | -555 (5.80/7.50) | -530 (6.13/7.96), *a*=2.947, *c*=12.950 |
| PZ, noSOC, Cr PP with 6 valence electrons, Hubbard $\mathbf{U_{eff}=1\ eV}$ at Cr-3d | | |
| AF | -184 (0.0/5.86) | -0.02 (0.0/4.16), *a*=2.853, *c*=12.711 |
| FM | -169 (3.48/4.33) | -42 (3.33/4.01), *a*=2.843, *c*=12.742 |
| PBE, SOC, Cr PP with 14 valence electrons | | |
| AF | Converge to PM | |
| FM | -116 (3.43/4.22) | -9 (3.14/3.72), *a*=2.851, *c*=12.776 |
| PBE, noSOC, Cr PP with 14 valence electrons | | |
| AF | Converges to PM | |
| FM | -117 (3.43/3.23) | -8 (3.14/3.72), *a*=2.851, *c*=12.774 |

## S3. Quantification of low temperature paramagnetism

To probe the origin of the low-temperature paramagnetic, PM, component in magnetization of $Cr_2AlC$, $M_{PM}$, and to determine its magnitude, we apply differential Brillouin function analysis, DBFA [7-11]. This approach is particularly suitable when a (weak) PM signal must be extracted from a (dominant) magnetic background whose magnetic field, *H*, dependence is unknown, but whose temperature, *T*, dependence is, or can be regarded, negligible at very low temperatures, that is where PM shows very strong *T*-dependence. The DBFA is especially effective in systems with coexisting diamagnetic, Pauli-like, and/or ferromagnetic/antiferromagnetic/blocked-superparamagnetic contributions.

*Data and preprocessing***.** We measure isothermal magnetization loops *M*(*H*) as functions of the magnetic field, *H*, at 2, 5, and 8 K (and a 300 K reference). To suppress the *T*-independent background, each low-*T* curve can be baseline-corrected by subtracting the *M*(*H*, 300 K) trace. A mild over-compensation seen in Fig. S3 is inconsequential for DBFA because the analysis uses pairwise differences between two nearby low-*T* curves, in which any residual *T*-independent part cancels.

The DBFA relies on the key observation that, at sufficiently low temperatures, the temperature dependence of non-paramagnetic contributions becomes negligible, whereas the magnetization of localized paramagnetic centers retains a strong temperature variability. Consequently, by subtracting isothermal magnetization curves measured at two nearby temperatures, the paramagnetic component can be isolated as:

$$\Delta M_{DBFA}(H) = M(T_1, H) - M(T_2, H) \approx M_{\mathrm{PM}}(T_1, H) - M_{\mathrm{PM}}(T_2, H),$$

where $M_{\mathrm{PM}}$ denotes the component brought about by the paramagnetic spins at two nearby temperatures $T_1$ and $T_2$, $T_1 < T_2$.

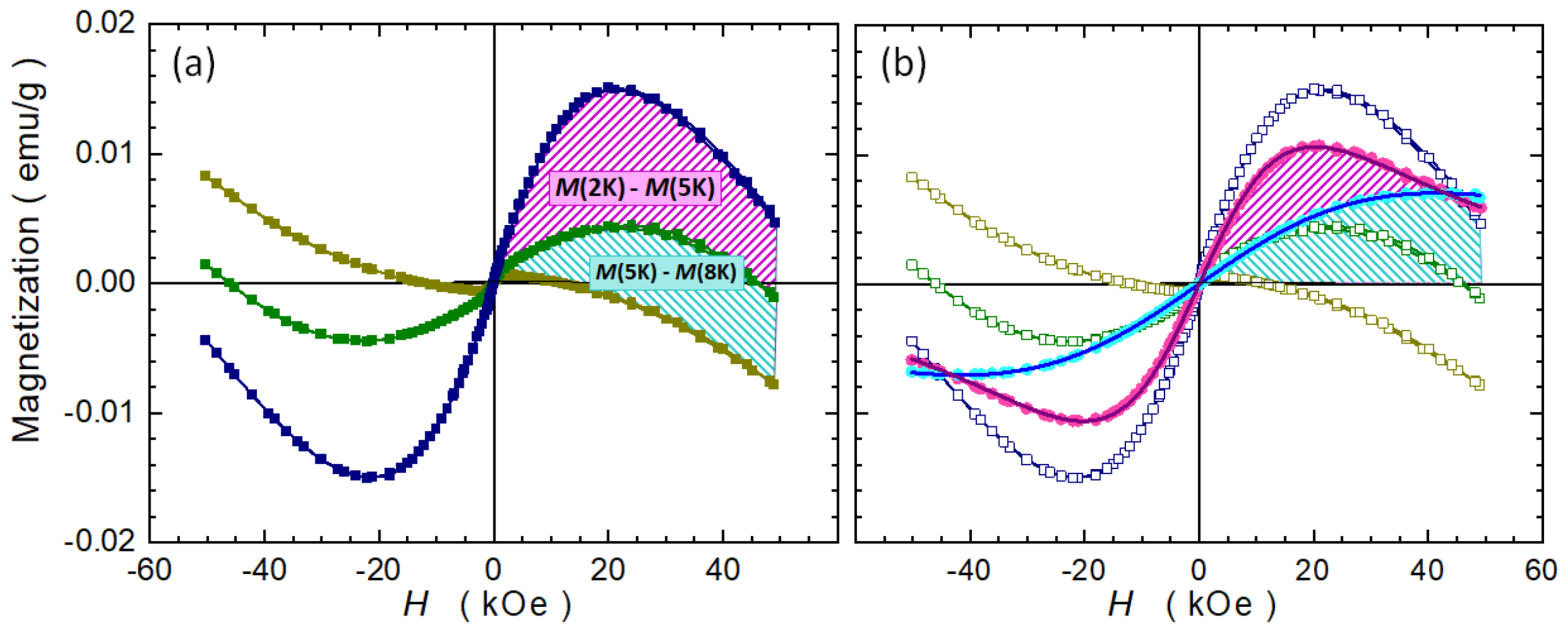

**Figure S3.** (**a**) Magnetic isotherms *M*(*H*) at low temperatures (indicated in the panel). The hatched area mark $\Delta M_{DBFA}(H)$, the differences between *M*(*H*) at 2 and 5 K and 5 and 8 K. (**b**) The same differences re-plotted on the same coordinate frame (bullets) with an equivalent difference between Brillouin functions $\Delta M_{teo}(H)$ calculated for spin $S = 3/2$ (solid lines of corresponding hues). The matching factor between $\Delta M_{DBFA}(H)$ and $\Delta M_{teo}(H)$, $n_S$, yields the concentration of paramagnetic spins in the sample.

These differential magnetization curves $\Delta M_{DBFA}(H)$, [bullets in Fig. S3(b)] can then be modeled using the difference of Brillouin functions taken at the same pairs of temperatures.

$$\Delta M_{teo}(H) = n_S g \mu_B S[B_S(T_1, H) - B_S(T_2, H)],$$

where $g$ is the Landé factor (taken here as $g = 2$, $\mu_B$ is the Bohr magneton, $n_S$ is the concentration of spins, and $B_S$ is the Brillouin function defined as:

$$B_S(x) = \frac{2S+1}{2S} coth\left(\frac{2S+1}{2S}x\right) - \frac{1}{2S} coth\left(\frac{1}{2S}x\right), \quad x = S\frac{g\mu_B H}{k_B T},$$

$k_B$ is the Boltzmann constant.

Importantly, the $\Delta M_{teo}(H)$, the difference between two Brillouin functions taken at nearby temperatures produces a characteristic maximum, the position of which – for a given differences of temperatures – is highly sensitive to the spin quantum number $S$. This makes the differential method exceptionally robust in determining the spin of the magnetic centers, even when the absolute PM signal is weak.

Fitting relevant $\Delta M_{teo}(H)$ to $\Delta M_{DBFA}(H)$ [the differences between $M(H)$ taken here at 2 and 5 K and 5 and 8 K – these differences are marked as the hatched areas in both panels of Fig. S3] yields $S \approx$ 1.45–1.55, indicating that the dominant spin component responsible for the low temperature paramagnetism comes from $S = 3/2$ spins. Final fits, thick solid lines in Fig. S3(b), performed with fixed $S = 3/2$ provide a very good agreement with experimental data, yielding $n_S \cong 1.5\times10^{18}$ spins per gram of $Cr_2AlC$, and so a very low fraction of paramagnetic spins $f \approx 0.00018$ (~0.02%).

Once the spin quantum number $S = 3/2$ and the concentration of localized paramagnetic centers $n_S$ have been determined by DBFA, their contribution to the temperature-dependent magnetization can be calculated over the entire experimental temperature range using the Brillouin function:

$$M_{PM}(T,H) = n_S g \mu_B S B_S\,(g\mu_B S \cdot H / k_B T).$$

Here, $M_{PM}$ is expressed per unit mass of the sample, consistently with $n_S$ given in spins per gram. For the temperature-dependent measurements performed at a constant magnetic field $H = 10$ kOe, the corresponding paramagnetic susceptibility is obtained as:

$$\chi_{PM}(T) = M_{PM}(T, H = 10 \text{ kOe}) / H.$$

Importantly, the full Brillouin function is used rather than the Curie-law approximation, since the magnetization of localized spins becomes nonlinear in magnetic field at sufficiently low temperatures. Thus, the independently determined parameters $S$, $g$, and $n_S$ allow us to calculate $\chi_{PM}(T)$ at every temperature without performing additional $M(H)$ measurements at each temperature point. This procedure assumes that the localized centers behave as independent paramagnetic spins with temperature-independent spin concentration and $g$-factor.

The calculated paramagnetic contribution is subsequently subtracted from the experimental susceptibility:

$\chi$corr(T) = $\chi$exp(T) − $\chi_{PM}(T)$.

As shown in Fig. S4, this correction effectively removes the low-temperature Curie-like upturn, and the resulting susceptibility becomes nearly temperature-independent below approximately 70 K, down to about 7 - 8 K. A small residual deviation at lower temperatures may indicate an additional, less abundant population of localized spins, which also contributes to the uncertainty in the determination of *S*. Nevertheless, the DBFA-based correction provides a reliable basis for analyzing the remaining temperature-dependent susceptibility in terms of antiferromagnetically coupled Cr-Cr pairs, as discussed in the main text.

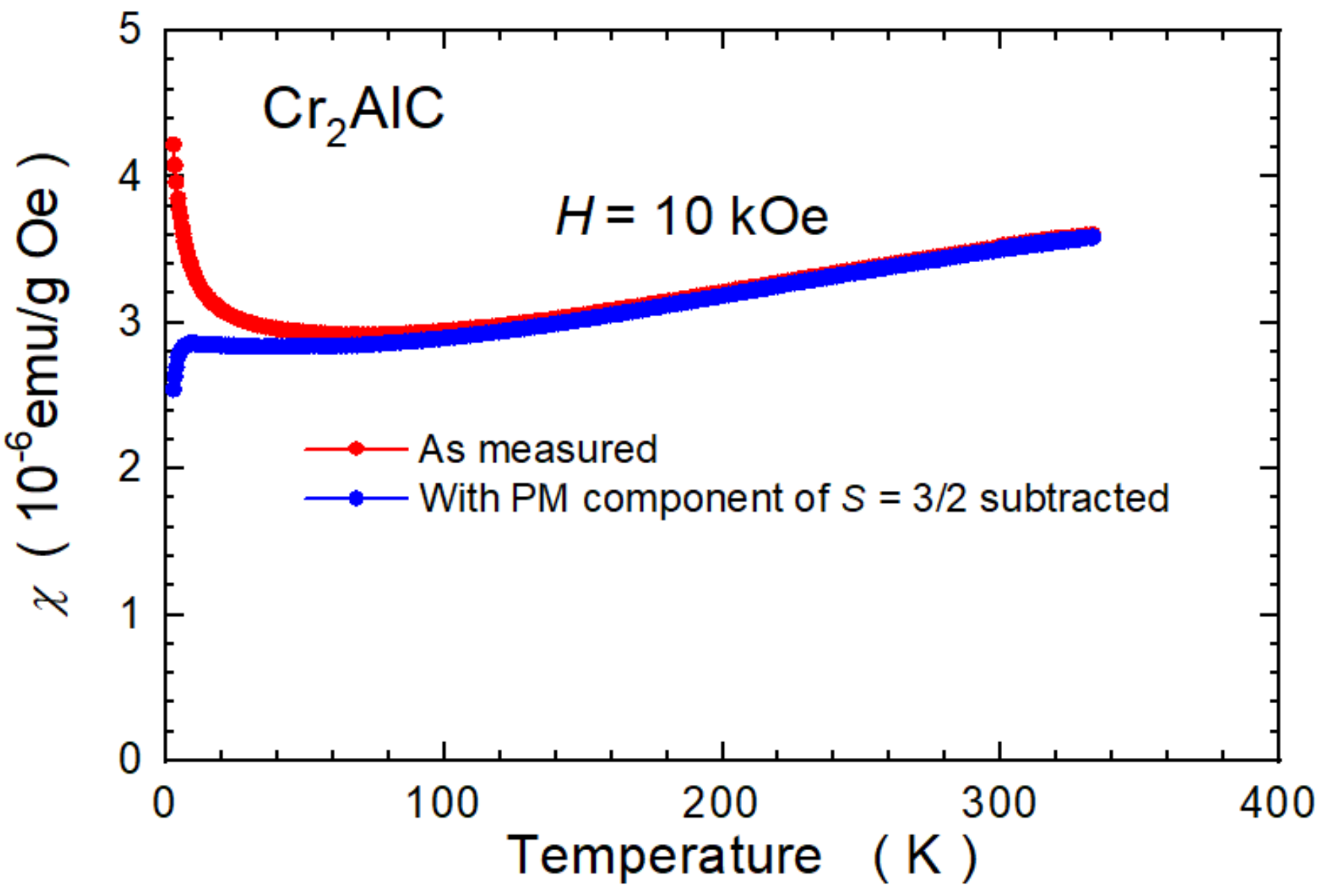

**Figure S4.** Temperature dependence of the magnetic susceptibility of $Cr_2AlC$ measured at H = 10 kOe before (red symbols) and after (blue symbols) subtraction of the localized paramagnetic contribution $\chi_{PM}(T)$. The latter was calculated using the full Brillouin function for $S$ = 3/2, $g$ = 2, and the spin concentration determined independently from the differential Brillouin function analysis (DBFA) of the M(H) curves measured at 2, 5, and 8 K.

## S4. Band structure calculated with the DFT

**Table S2.** List of the high symmetry points in the Brillouin zone (BZ) for the hexagonal crystal structure with the space group $P6_3/mmc$. The parameters *a* and *c* are the basal and vertical lattice parameters, respectively.

| *k*-point | (u, v, w) | ($k_x$, $k_y$, $k_z$) |
|---|---|---|
| Γ | (0, 0, 0) | [0, 0, 0] |
| M | (1/2, 0, 0) | [$\pi/a$, $-\pi/\sqrt{3}a$, 0] |
| K | (2/3, 1/3, 0) | [$4\pi/3a$, 0, 0] |
| A | (0, 0, 1/2) | [0, 0, $\pi/c$] |
| L | (½, 0, 1/2) | [$\pi/a$, $-\pi/\sqrt{3}a$, $\pi/c$] |
| H | (2/3, 1/3, 1/2) | [$4\pi/3a$, 0, $\pi/c$] |

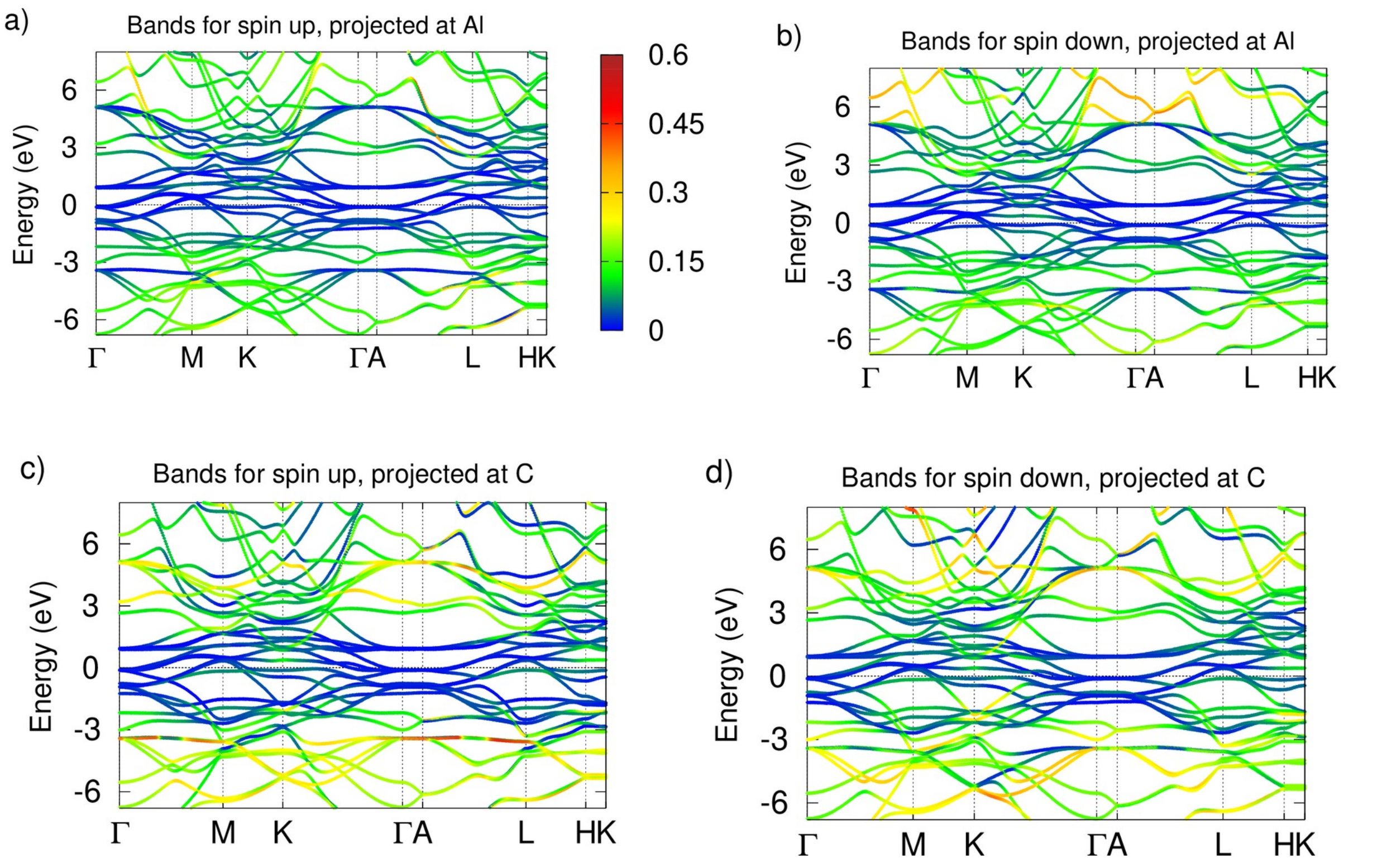

**Figure S5.** The spin-bands projected at the maximally-localized Wannier functions centered at: a) Al atoms for spin up, b) Al atoms for spin down, c) C atoms for spin up and d) C atoms for spin down. Color legend is common for all cases and truncated to the [0.0,0.6] range for a better contrast.

**S5. Absorption spectra obtained with the *ai*-BSE**

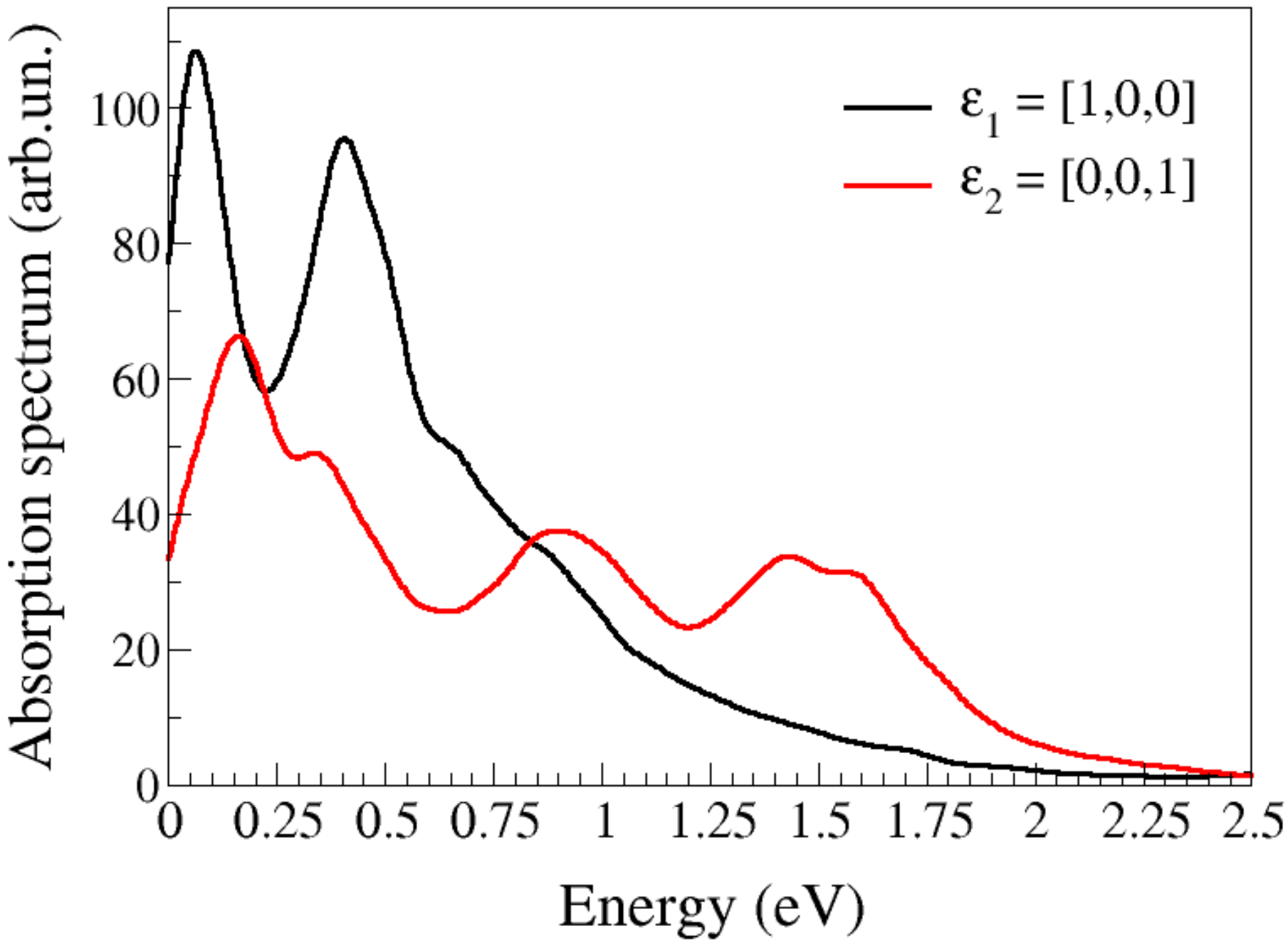

**Figure S6.** Comparison of the not normalized absorption spectra obtained from the solution of *ai*-BSE with the electric field vector oriented along the [1,0,0] and [0,0,1] crystal directions.

**Table S3.** The list of *k*-points in the irreducible symmetry wedge of BZ (IBZ) used for solving the *ab initio* BSE with the use of the uniform mesh of the 12×12×4 grid.

| No. | Cartesian coordinates in units 2π/*a* (*a* is the lattice constant = 5.4983 a.u.) | | | Crystal coordinates | | |
|---|---|---|---|---|---|---|
| 1 | 0.0000000 | 0.0000000 | 0.0000000 | 0.0000000 | 0.0000000 | 0.0000000 |
| 2 | 0.0000000 | 0.0000000 | 0.0562985 | 0.0000000 | 0.0000000 | 0.2500000 |
| 3 | 0.0000000 | 0.0000000 | -0.1125971 | 0.0000000 | 0.0000000 | -0.5000000 |
| 4 | 0.0000000 | 0.0963226 | 0.0000000 | 0.0000000 | 0.0833333 | 0.0000000 |
| 5 | 0.0000000 | 0.0963226 | 0.0562985 | 0.0000000 | 0.0833333 | 0.2500000 |
| 6 | 0.0000000 | 0.0963226 | -0.1125971 | 0.0000000 | 0.0833333 | -0.5000000 |
| 7 | 0.0000000 | 0.1926452 | 0.0000000 | 0.0000000 | 0.1666667 | 0.0000000 |
| 8 | 0.0000000 | 0.1926452 | 0.0562985 | 0.0000000 | 0.1666667 | 0.2500000 |
| 9 | 0.0000000 | 0.1926452 | -0.1125971 | 0.0000000 | 0.1666667 | -0.5000000 |
| 10 | 0.0000000 | 0.2889678 | 0.0000000 | 0.0000000 | 0.2500000 | 0.0000000 |
| 11 | 0.0000000 | 0.2889678 | 0.0562985 | 0.0000000 | 0.2500000 | 0.2500000 |
| 12 | 0.0000000 | 0.2889678 | -0.1125971 | 0.0000000 | 0.2500000 | -0.5000000 |
| 13 | 0.0000000 | 0.3852905 | 0.0000000 | 0.0000000 | 0.3333333 | 0.0000000 |
| 14 | 0.0000000 | 0.3852905 | 0.0562985 | 0.0000000 | 0.3333333 | 0.2500000 |
| 15 | 0.0000000 | 0.3852905 | -0.1125971 | 0.0000000 | 0.3333333 | -0.5000000 |
| 16 | 0.0000000 | 0.4816131 | 0.0000000 | 0.0000000 | 0.4166667 | 0.0000000 |
| 17 | 0.0000000 | 0.4816131 | 0.0562985 | 0.0000000 | 0.4166667 | 0.2500000 |
| 18 | 0.0000000 | 0.4816131 | -0.1125971 | 0.0000000 | 0.4166667 | -0.5000000 |
| 19 | 0.0000000 | -0.5779357 | 0.0000000 | 0.0000000 | -0.5000000 | 0.0000000 |
| 20 | 0.0000000 | -0.5779357 | 0.0562985 | 0.0000000 | -0.5000000 | 0.2500000 |
| 21 | 0.0000000 | -0.5779357 | -0.1125971 | 0.0000000 | -0.5000000 | -0.5000000 |
| 22 | 0.0834178 | 0.1444839 | 0.0000000 | 0.0833333 | 0.0833333 | 0.0000000 |
| 23 | 0.0834178 | 0.1444839 | 0.0562985 | 0.0833333 | 0.0833333 | 0.2500000 |
| 24 | 0.0834178 | 0.1444839 | -0.1125971 | 0.0833333 | 0.0833333 | -0.5000000 |
| 25 | 0.0834178 | 0.2408065 | 0.0000000 | 0.0833333 | 0.1666667 | 0.0000000 |
| 26 | 0.0834178 | 0.2408065 | 0.0562985 | 0.0833333 | 0.1666667 | 0.2500000 |
| 27 | 0.0834178 | 0.2408065 | -0.1125971 | 0.0833333 | 0.1666667 | -0.5000000 |
| 28 | 0.0834178 | 0.3371292 | 0.0000000 | 0.0833333 | 0.2500000 | 0.0000000 |
| 29 | 0.0834178 | 0.3371292 | 0.0562985 | 0.0833333 | 0.2500000 | 0.2500000 |
| 30 | 0.0834178 | 0.3371292 | -0.1125971 | 0.0833333 | 0.2500000 | -0.5000000 |
| 31 | 0.0834178 | 0.4334518 | 0.0000000 | 0.0833333 | 0.3333333 | 0.0000000 |
| 32 | 0.0834178 | 0.4334518 | 0.0562985 | 0.0833333 | 0.3333333 | 0.2500000 |
| 33 | 0.0834178 | 0.4334518 | -0.1125971 | 0.0833333 | 0.3333333 | -0.5000000 |
| 34 | 0.0834178 | 0.5297744 | 0.0000000 | 0.0833333 | 0.4166667 | 0.0000000 |
| 35 | 0.0834178 | 0.5297744 | 0.0562985 | 0.0833333 | 0.4166667 | 0.2500000 |
| 36 | 0.0834178 | 0.5297744 | -0.1125971 | 0.0833333 | 0.4166667 | -0.5000000 |
| 37 | 0.1668357 | 0.2889678 | 0.0000000 | 0.1666667 | 0.1666667 | 0.0000000 |
| 38 | 0.1668357 | 0.2889678 | 0.0562985 | 0.1666667 | 0.1666667 | 0.2500000 |
| 39 | 0.1668357 | 0.2889678 | -0.1125971 | 0.1666667 | 0.1666667 | -0.5000000 |
| 40 | 0.1668357 | 0.3852905 | 0.0000000 | 0.1666667 | 0.2500000 | 0.0000000 |
| 41 | 0.1668357 | 0.3852905 | 0.0562985 | 0.1666667 | 0.2500000 | 0.2500000 |
| 42 | 0.1668357 | 0.3852905 | -0.1125971 | 0.1666667 | 0.2500000 | -0.5000000 |
| 43 | 0.1668357 | 0.4816131 | 0.0000000 | 0.1666667 | 0.3333333 | 0.0000000 |
| 44 | 0.1668357 | 0.4816131 | 0.0562985 | 0.1666667 | 0.3333333 | 0.2500000 |
| 45 | 0.1668357 | 0.4816131 | -0.1125971 | 0.1666667 | 0.3333333 | -0.5000000 |
| 46 | 0.1668357 | 0.5779357 | 0.0000000 | 0.1666667 | 0.4166667 | 0.0000000 |
| 47 | 0.1668357 | 0.5779357 | 0.0562985 | 0.1666667 | 0.4166667 | 0.2500000 |
| 48 | 0.1668357 | 0.5779357 | -0.1125971 | 0.1666667 | 0.4166667 | -0.5000000 |
| 49 | 0.2502535 | 0.4334518 | 0.0000000 | 0.2500000 | 0.2500000 | 0.0000000 |
| 50 | 0.2502535 | 0.4334518 | 0.0562985 | 0.2500000 | 0.2500000 | 0.2500000 |
| 51 | 0.2502535 | 0.4334518 | -0.1125971 | 0.2500000 | 0.2500000 | -0.5000000 |
| 52 | 0.2502535 | 0.5297744 | 0.0000000 | 0.2500000 | 0.3333333 | 0.0000000 |
| 53 | 0.2502535 | 0.5297744 | 0.0562985 | 0.2500000 | 0.3333333 | 0.2500000 |
| 54 | 0.2502535 | 0.5297744 | -0.1125971 | 0.2500000 | 0.3333333 | -0.5000000 |
| 55 | 0.3336713 | 0.5779357 | 0.0000000 | 0.3333333 | 0.3333333 | 0.0000000 |
| 56 | 0.3336713 | 0.5779357 | 0.0562985 | 0.3333333 | 0.3333333 | 0.2500000 |
| 57 | 0.3336713 | 0.5779357 | -0.1125971 | 0.3333333 | 0.3333333 | -0.5000000 |
| 58 | 0.0000000 | -0.0963226 | 0.0562985 | 0.0000000 | -0.0833333 | 0.2500000 |
| 59 | 0.0000000 | -0.1926452 | 0.0562985 | 0.0000000 | -0.1666667 | 0.2500000 |
| 60 | 0.0000000 | -0.2889678 | 0.0562985 | 0.0000000 | -0.2500000 | 0.2500000 |
| 61 | 0.0000000 | -0.3852905 | 0.0562985 | 0.0000000 | -0.3333333 | 0.2500000 |
| 62 | 0.0000000 | -0.4816131 | 0.0562985 | 0.0000000 | -0.4166667 | 0.2500000 |
| 63 | -0.0834178 | -0.2408065 | 0.0562985 | -0.0833333 | -0.1666667 | 0.2500000 |
| 64 | -0.0834178 | -0.3371292 | 0.0562985 | -0.0833333 | -0.2500000 | 0.2500000 |
| 65 | -0.0834178 | -0.4334518 | 0.0562985 | -0.0833333 | -0.3333333 | 0.2500000 |
| 66 | -0.0834178 | -0.5297744 | 0.0562985 | -0.0833333 | -0.4166667 | 0.2500000 |
| 67 | -0.1668357 | -0.3852905 | 0.0562985 | -0.1666667 | -0.2500000 | 0.2500000 |
| 68 | -0.1668357 | -0.4816131 | 0.0562985 | -0.1666667 | -0.3333333 | 0.2500000 |
| 69 | -0.2502535 | -0.5297744 | 0.0562985 | -0.2500000 | -0.3333333 | 0.2500000 |

**Table S4.** The list of excitons and contributing dipole transitions (i.e. band-to-band excitations at given *k*-point in BZ): energy of the exciton ($E_X$), oscillator strength (*f*), number of the given excitation in the whole spectrum, dominant excitations from the occupied band ($b_{val}$) to the unoccupied band ($b_{cond}$) at the corresponding *k*-point in BZ (according to the list in Table S2) and the spin channel (up denored "U" and/or down denoted "D") that causes a change of the local magnetization (ΔM). The weights of the listed band-to-band contributions are larger than 5%.

| $E_X$ (eV) | *f* | No. | $b_{val}$ | $b_{cond}$ | *k* | spin | ΔM |
|---|---|---|---|---|---|---|---|
| **Electric field vector [1,0,0]** | | | | | | | |
| 0.03896028 | 0.4337 | 255 | | | | | |
| 0.03896082 | 0.7144 | 256 | | | | | |
| 0.03900557 | 1.0000 | 259 | 45 | 46 | 9 | U D | -0.113 |
| 0.03900839 | 0.6842 | 260 | 42 | 43 | 18 | U D | -0.026 |
| 0.03954767 | 0.4273 | 268 | | | | | |
| 0.03955289 | 0.8781 | 269 | | | | | |
| 0.03960037 | 0.7446 | 271 | | | | | |
| 0.03960456 | 0.7899 | 272 | | | | | |
| 0.04272138 | 0.2391 | 287 | 45 | 46 | 5 | U D | -0.036 |
| 0.04273618 | 0.1863 | 289 | 45 | 46 | 58 | U D | -0.069 |
| 0.04310656 | 0.2313 | 298 | | | | | |
| 0.04910249 | 0.3118 | 334 | | | | | |
| 0.04916531 | 0.3645 | 337 | | | | | |
| 0.04916605 | 0.2138 | 338 | 44 | 46 | 9 | U D | -0.105 |
| 0.04996306 | 0.1986 | 340 | | | | | |
| 0.04996548 | 0.2612 | 341 | | | | | |
| 0.05002971 | 0.4124 | 343 | | | | | |
| 0.05003283 | 0.1919 | 344 | | | | | |
| 0.06341520 | 0.2312 | 406 | 45 | 46 | 4 | U D | -0.044 |
| 0.06344353 | 0.1774 | 410 | 42 | 43 | 16 | U D | -0.035 |
| 0.06388643 | 0.2000 | 413 | | | | | |
| 0.08932626 | 0.1310 | 590 | | | | | |
| 0.08934113 | 0.5461 | 591 | 44 | 45 | 23 | U D | -0.083 |
| 0.08938697 | 0.4277 | 600 | 45 | 46 | 8 | U D | -0.139 |
| 0.08939309 | 0.1330 | 601 | 45 | 46 | 59 | U D | -0.158 |
| 0.08979803 | 0.6369 | 608 | | | | | |
| 0.08999798 | 0.2048 | 610 | | | | | |
| 0.09000240 | 0.7790 | 611 | | | | | |
| 0.09008569 | 0.4336 | 614 | | | | | |
| 0.09042332 | 0.1561 | 619 | | | | | |
| 0.09042566 | 0.2242 | 620 | | | | | |
| 0.13229594 | 0.1015 | 898 | | | | | |
| 0.13233391 | 0.2847 | 901 | 45 | 46 | 7 | U D | -0.170 |
| 0.13295464 | 0.2160 | 904 | | | | | |
| 0.13300720 | 0.2131 | 907 | | | | | |
| 0.21025726 | 0.1442 | 1508 | 41 | 42 | 35 | U D | -0.244 |
| 0.21025996 | 0.1283 | 1509 | 41 | 42 | 66 | U D | -0.252 |

| | | | | | | | | |
|---|---|---|---|---|---|---|---|---|
| 0.21130331 | 0.1656 | 1547 | 41<br>41<br>42<br>42<br>45 | 43<br>43<br>43<br>43<br>46 | 61<br>14<br>29<br>64<br>23 | U D<br>U D<br>U<br>D<br>U D | 0.242<br>0.268<br>-0.079<br>0.003<br>0.001 |
| 0.27909166<br>0.27909717<br>0.28036043<br>0.28036541<br>0.28110805 | 0.1916<br>0.2031<br>0.2299<br>0.1396<br>0.2157 | 1962<br>1963<br>1975<br>1976<br>1984 | 41<br>43 | 43<br>45 | 36<br>24 | U D<br>U D | -0.262<br>-0.234 |
| 0.33812738<br>0.33813211<br>0.33960593<br>0.34089473 | 0.1149<br>0.1349<br>0.3685<br>0.2452 | 2418<br>2419<br>2458<br>2491 | 41<br>41<br>41<br>41<br>41<br>41<br>43<br>43 | 42<br>42<br>42<br>42<br>43<br>43<br>44<br>46 | 16<br>17<br>18<br>62<br>35<br>66<br>40<br>9 | U<br>U<br>U<br>U<br>U<br>U<br>U D<br>U D | 0.127<br>0.127<br>0.135<br>0.130<br>-0.238<br>-0.208<br>0.189<br>-0.043 |
| 0.36765716 | 0.2302 | 2823 | 42<br>42<br>43 | 43<br>43<br>45 | 58<br>69<br>39 | U D<br>D<br>U D | -0.007<br>0.004<br>-0.341 |
| 0.37915465<br>0.37915918 | 0.1401<br>0.1268 | 2971<br>2972 | 41<br>41 | 43<br>43 | 33<br>34 | U<br>D | -0.020<br>0.052 |
| 0.39388055<br>0.39738923 | 0.2072<br>0.1145 | 3218<br>3291 | 41<br>41<br>41<br>41<br>43 | 43<br>43<br>43<br>43<br>44 | 17<br>32<br>62<br>65<br>42 | U D<br>U D<br>U D<br>U D<br>D | 0.140<br>0.006<br>0.154<br>0.016<br>0.001 |
| 0.41265944<br>0.41288036<br>0.41288352<br>0.41328147<br>0.41444153<br>0.41932255<br>0.41984427 | 0.1341<br>0.1870<br>0.2219<br>0.3209<br>0.1021<br>0.3274<br>0.2163 | 3514<br>3517<br>3518<br>3535<br>3590<br>3650<br>3658 | 41<br>42<br>43<br>43<br>43<br>43<br>43 | 43<br>44<br>45<br>45<br>45<br>45<br>45 | 31<br>18<br>31<br>32<br>33<br>38<br>65 | U<br>U D<br>U D<br>U D<br>U D<br>U D<br>U D | 0.021<br>-0.225<br>0.009<br>-0.027<br>-0.042<br>-0.237<br>-0.078 |
| 0.42449072<br>0.42524788 | 0.2174<br>0.1798 | 3718<br>3732 | 41<br>43 | 43<br>45 | 47<br>42 | D<br>U D | 0.047<br>0.037 |
| 0.44782135<br>0.44914323 | 0.1441<br>0.1549 | 3922<br>3948 | 43<br>43<br>45 | 45<br>45<br>46 | 41<br>67<br>39 | U D<br>U D<br>U D | 0.340<br>0.222<br>-0.215 |
| 0.49753270<br>0.49753615<br>0.49760446<br>0.49760595<br>0.49779108 | 0.1904<br>0.2023<br>0.1772<br>0.1791<br>0.4011 | 4508<br>4509<br>4511<br>4512<br>4520 | 42<br>43 | 44<br>44 | 31<br>48 | U D<br>U | 0.010<br>0.156 |
| 0.50783712<br>0.50783974 | 0.2000<br>0.1584 | 4725<br>4726 | 42<br>42 | 43<br>44 | 32<br>32 | D<br>U | 0.007<br>-0.032 |
| 0.50799561<br>0.50799865 | 0.1596<br>0.2578 | 4732<br>4733 | 42<br>43<br>43 | 44<br>44<br>44 | 53<br>65<br>69 | U D<br>U D<br>U D | -0.064<br>-0.044<br>-0.140 |
| 0.61652535 | 0.1460 | 5743 | 41<br>42<br>42<br>42 | 44<br>45<br>45<br>45 | 20<br>22<br>23<br>37 | D<br>D<br>D<br>U D | 0.002<br>0.008<br>0.007<br>-0.153 |
| 0.62833184<br>0.62869668 | 0.1873<br>0.2761 | 5839<br>5846 | 42 | 44 | 40 | U D | 0.373 |
| 0.65360451 | 0.1114 | 6091 | 42<br>42 | 44<br>45 | 47<br>25 | D<br>U D | 0.052<br>-0.425 |
| 0.68262810<br>0.68262893 | 0.1722<br>0.1089 | 6536<br>6537 | 42<br>42<br>42<br>43 | 44<br>44<br>44<br>44 | 41<br>61<br>67<br>14 | U D<br>D<br>U D<br>D | 0.345<br>0.038<br>0.304<br>0.002 |
| | | | | | | | |
| $E_X$ (eV) | f | No. | $b_{val}$ | $b_{cond}$ | k | spin | ΔM |
| **Electric field vector [0,0,1]** | | | | | | | |
| 0.00997620 | 0.2071 | 18 | 44 | 45 | 9 | U D | 0.008 |
| 0.03551616 | 1.0000 | 236 | 42 | 43 | 36 | U D | -0.038 |
| 0.10666842<br>0.10667458 | 0.9232<br>0.3588 | 701<br>703 | 45 | 46 | 24 | U D | 0.035 |
| 0.12827328 | 0.1003 | 849 | 42<br>42 | 43<br>43 | 35<br>66 | U<br>D | 0.006<br>0.003 |
| 0.13607901 | 0.2495 | 991 | 42<br>42<br>44 | 43<br>43<br>45 | 33<br>46<br>38 | U D<br>U D<br>U D | 0.046<br>0.063<br>-0.164 |
| 0.15361083<br>0.16569342 | 0.6429<br>0.2490 | 1123<br>1195 | 41<br>42<br>43<br>44<br>44 | 42<br>43<br>45<br>45<br>46 | 61<br>47<br>4<br>39<br>7 | U<br>U D<br>U<br>D<br>D | 0.254<br>0.242<br>-0.089<br>0.085<br>0.017 |
| 0.17703396 | 0.1202 | 1233 | 43<br>43<br>44 | 46<br>46<br>45 | 5<br>58<br>26 | U D<br>U D<br>U | -0.063<br>-0.050<br>-0.030 |
| 0.17849523<br>0.17917451 | 0.5071<br>0.1151 | 1263<br>1277 | 44<br>44 | 45<br>45 | 26<br>63 | U D<br>U D | -0.077<br>-0.079 |
| 0.20491204 | 0.5662 | 1412 | 43 | 44 | 39 | U D | -0.210 |
| 0.21571817 | 0.1004 | 1589 | 44 | 45 | 27 | U D | -0.105 |
| 0.21644777 | 0.2235 | 1603 | 42 | 43 | 45 | U D | 0.319 |
| 0.27991608 | 0.1147 | 1987 | 43 | 45 | 24 | U D | -0.234 |
| 0.33588421 | 0.5619 | 2420 | 41<br>41<br>41<br>41<br>41<br>43 | 42<br>42<br>42<br>43<br>43<br>44 | 16<br>17<br>62<br>35<br>66<br>40 | D<br>U D<br>U D<br>U D<br>U D<br>D | 0.001<br>0.128<br>0.130<br>-0.177<br>-0.208<br>-0.014 |
| 0.33765310<br>0.33781430<br>0.33867326 | 0.1616<br>0.1855<br>0.2200 | 2445<br>2456<br>2473 | 41<br>41<br>42<br>43 | 42<br>42<br>43<br>46 | 17<br>18<br>52<br>9 | U<br>U<br>U D<br>U D | 0.127<br>0.134<br>0.429<br>-0.043 |
| 0.36926982 | 0.1090 | 2903 | 43 | 44 | 43 | U D | 0.228 |
| 0.38676953 | 0.2653 | 3110 | 43 | 46 | 24 | U D | -0.199 |
| 0.39896798 | 0.1760 | 3378 | 43<br>43<br>43<br>43 | 45<br>45<br>45<br>45 | 8<br>17<br>59<br>62 | D<br>U<br>D<br>U | -0.053<br>-0.235<br>0.020<br>-0.276 |
| 0.43284208 | 0.2654 | 3824 | 42 | 45 | 36 | U D | 0.445 |
| 0.48671335 | 0.2483 | 4427 | 43<br>43<br>43 | 44<br>46<br>46 | 50<br>8<br>59 | U<br>U D<br>U D | 0.083<br>-0.115<br>-0.101 |
| 0.49706751 | 0.1047 | 4547 | 43<br>43<br>43<br>44<br>44 | 44<br>45<br>46<br>46<br>46 | 23<br>27<br>48<br>26<br>63 | U D<br>U D<br>U D<br>U D<br>U D | -0.223<br>-0.357<br>0.160<br>-0.118<br>-0.143 |
| 0.53209925 | 0.1153 | 5101 | 43<br>43 | 45<br>45 | 26<br>63 | U D<br>U D | -0.307<br>-0.383 |
| 0.63824880 | 0.1918 | 5938 | 41<br>43 | 44<br>46 | 19<br>7 | D<br>U D | 0.003<br>-0.235 |
| 0.72256321<br>0.72300303 | 0.1400<br>0.2358 | 7169<br>7179 | 43<br>43<br>43<br>43<br>42<br>42<br>42 | 46<br>46<br>46<br>46<br>44<br>44<br>44 | 12<br>17<br>27<br>62<br>13<br>14<br>61 | U D<br>U D<br>U D<br>U<br>U<br>U D<br>U | -0.186<br>-0.635<br>-0.262<br>-0.725<br>-0.373<br>-0.383<br>-0.446 |
| 0.81346548 | 0.2149 | 9188 | 41<br>41<br>41<br>41<br>43 | 45<br>45<br>45<br>46<br>46 | 3<br>32<br>65<br>3<br>39 | D<br>U<br>U<br>D<br>U | 0.055<br>-0.100<br>-0.128<br>0.035<br>-0.575 |
| 0.82699269 | 0.2954 | 9467 | 43<br>43 | 46<br>46 | 11<br>60 | U D<br>U | -0.351<br>-0.340 |
| 0.89028001<br>0.89047629 | 0.1651<br>0.1253 | 10594<br>10602 | 41<br>41<br>42 | 44<br>44<br>46 | 14<br>61<br>7 | D<br>D<br>D | -0.001<br>0.039<br>0.009 |
| | | | 43<br>45 | 46<br>47 | 42<br>24 | D<br>U D | -0.068<br>-0.015 |
| 0.93515009<br>0.93539619 | 0.1511<br>0.1274 | 11291<br>11296 | 43<br>43<br>44<br>44<br>44<br>44<br>44<br>44<br>45 | 46<br>46<br>47<br>47<br>47<br>48<br>48<br>48<br>47 | 38<br>64<br>8<br>9<br>59<br>7<br>8<br>59<br>22 | U D<br>U<br>U<br>D<br>U<br>U<br>D<br>U D<br>U D | -0.457<br>-0.663<br>-0.143<br>0.033<br>-0.183<br>-0.215<br>0.091<br>-0.107<br>-0.071 |
| 1.02420008 | 0.1471 | 12867 | 41<br>42<br>43<br>43<br>43<br>43<br>44<br>44<br>44<br>44<br>44<br>44<br>44 | 46<br>45<br>47<br>47<br>48<br>48<br>47<br>47<br>48<br>48<br>48<br>48<br>48 | 14<br>43<br>1<br>2<br>1<br>2<br>1<br>2<br>1<br>2<br>5<br>6<br>58 | U<br>U D<br>U<br>U<br>U D<br>U<br>U<br>U<br>U<br>U<br>D<br>U D<br>D | -0.607<br>0.474<br>-0.508<br>-0.514<br>-0.592<br>-0.603<br>-0.514<br>-0.507<br>-0.598<br>-0.597<br>0.080<br>-0.297<br>0.006 |
| 1.28471279 | 0.1060 | 15725 | 42 | 45 | 52 | U D | 0.412 |
| 1.36124873 | 0.1301 | 16698 | 43<br>43 | 48<br>48 | 35<br>66 | D<br>U D | 0.015<br>-0.473 |
| 1.37435663 | 0.1854 | 16851 | 43<br>43 | 48<br>48 | 41<br>67 | U D<br>U D | -0.267<br>-0.255 |
| 1.40735841 | 0.1035 | 17147 | 42<br>43 | 46<br>48 | 40<br>68 | U D<br>U | -0.084<br>-0.388 |
| 1.43207085 | 0.1257 | 17486 | 41 | 47 | 18 | U D | 0.095 |
| 1.46015453 | 0.1565 | 17779 | 42<br>43<br>43 | 47<br>47<br>47 | 11<br>55<br>60 | D<br>U<br>U | 0.039<br>0.182<br>-0.230 |
| 1.57289636 | 0.2369 | 18999 | 41 | 48 | 12 | U D | -0.764 |
| 1.59796524 | 0.1047 | 19310 | 41 | 48 | 15 | U | -0.696 |
| 1.60620856 | 0.1933 | 19347 | 41<br>41<br>42 | 47<br>47<br>48 | 32<br>65<br>27 | U D<br>U D<br>U D | -0.254<br>-0.269<br>-0.454 |
| 1.60795391 | 0.1042 | 19393 | 41<br>41 | 47<br>47 | 28<br>65 | U D<br>D | -0.494<br>0.121 |
| 1.79012108 | 0.1663 | 21338 | 41<br>42 | 48<br>48 | 6<br>52 | D<br>D | 0.119<br>0.149 |

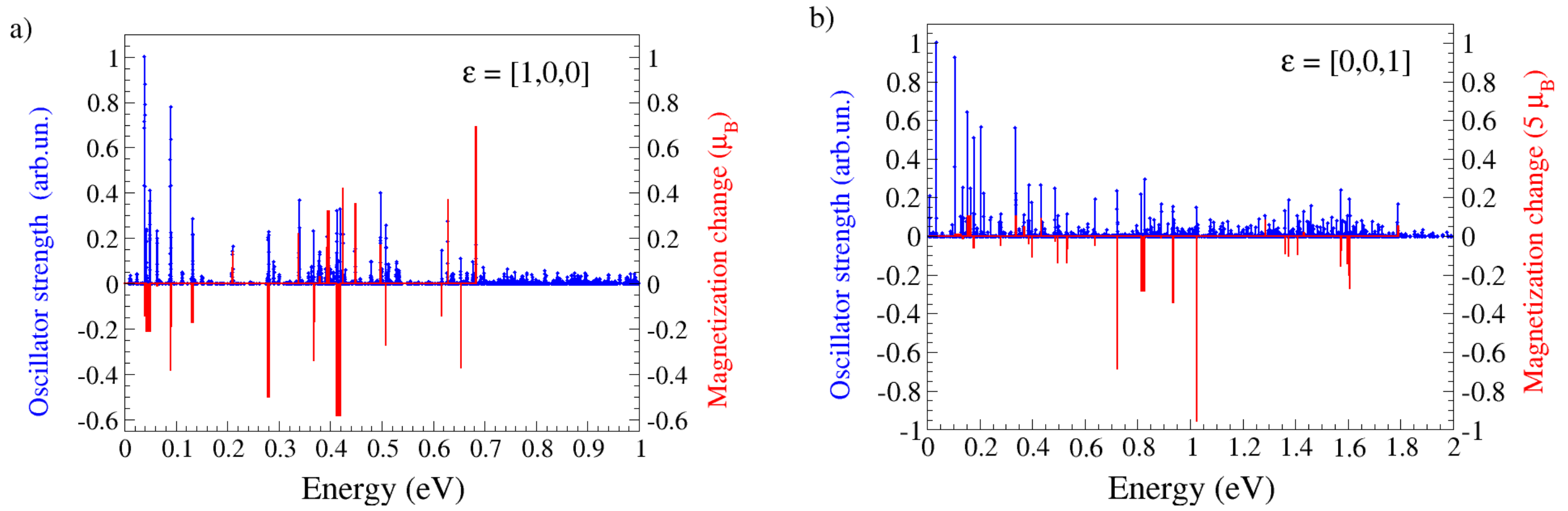

**Figure S7.** Normalized oscillator strengths (in blue color) of the dipole transitions in the absorption spectrum obtained from the solution of *ab initio* BSE with the electric-field vector oriented along: **a)** the [1,0,0] crystal direction and **b)** [0,0,1] crystal axis. The corresponding change of the local magnetization in the elementary cell (in red color) caused by the dominant excitations are given for the transitions with the oscillator strength threshold > 0.1. Note that the scale of the magnetization axis is 5 larger for [0,0,1] than for [1,0,0].